\documentclass[twocolumn]{aastex631}

\usepackage{natbib}
\usepackage{textcomp}

\shorttitle{BAL Quasars in GNIRS-DQS}
\shortauthors{Ahmed et al.}

\begin{document}

\title{Gemini Near Infrared Spectrograph - Distant Quasar Survey: Rest-Frame Ultraviolet-Optical Spectral Properties of Broad Absorption Line Quasars}

\author[0009-0008-0969-4084]{Harum Ahmed}
\affiliation {Department of Physics, University of North Texas, Denton, TX 76203, USA} \email{HarumAhmed@my.unt.edu}

\author[0000-0003-4327-1460]{Ohad Shemmer}
\affiliation{Department of Physics, University of North Texas, Denton, TX 76203, USA}

\author[0000-0001-8406-4084]{Brandon Matthews}
\affiliation{Department of Physics, University of North Texas, Denton, TX 76203, USA}

\author[0000-0003-0192-1840]{Cooper Dix}
\affiliation{Department of Physics, University of North Texas, Denton, TX 76203, USA}
\affiliation{Department of Astronomy, The University of Texas at Austin, Austin, TX 78712, USA}

\author[0000-0001-6600-2517]{Trung Ha}
\affiliation{Department of Physics, University of North Texas, Denton, TX 76203, USA}
\affiliation{Center for Computational Astrophysics, Flatiron Institute, 162 Fifth Avenue, New York, NY 10010, USA}

\author[0000-0002-1061-1804]{Gordon T. Richards}
\affiliation{Department of Physics, Drexel University, 32 S. 32nd Street, Philadelphia, PA 19104, USA}

\author[0000-0002-1207-0909]{Michael S. Brotherton}
\affiliation{Department of Physics and Astronomy, University of Wyoming, Laramie, WY 82071, USA}

\author{Adam D. Myers}
\affiliation{Department of Physics and Astronomy, University of Wyoming, Laramie, WY 82071, USA}

\author[0000-0002-0167-2453]{W. N. Brandt}
\affiliation{Department of Astronomy and Astrophysics, The Pennsylvania State University, University Park, PA 16802, USA} 
\affiliation{Institute for Gravitation and the Cosmos, The Pennsylvania State University, University Park, PA 16802, USA} 
\affiliation{Department of Physics, 104 Davey Lab, The Pennsylvania State University, University Park, PA 16802, USA}

\author[0000-0001-6217-8101]{Sarah C. Gallagher}
\affiliation{Department of Physics \& Astronomy, University of Western Ontario, 1151 Richmond St, London, ON N6C 1T7, Canada} 

\author[0000-0003-1245-5232]{Richard Green}
\affiliation{Steward Observatory, University of Arizona, 933 N Cherry Ave, Tucson, AZ 85721, USA} 

\author[0000-0003-1523-9164]{Paulina Lira}
\affiliation{Departmento de Astronom{\`i}a, Universidad de Chile, Casilla 36D, Santiago, Chile} 

\author[0000-0003-1081-2929]{Jacob N. McLane}
\affiliation{Department of Physics and Astronomy, University of Wyoming, Laramie, WY 82071, USA} 

\author[0000-0002-7092-0326]{Richard M. Plotkin}
\affiliation{Department of Physics, University of Nevada, Reno, NV 89557, USA} 
\affiliation{Nevada Center for Astrophysics, University of Nevada, Las Vegas, NV 89154, USA}

\author[0000-0001-7240-7449]{Donald P. Schneider}
\affiliation{Department of Astronomy and Astrophysics, The Pennsylvania State University, University Park, PA 16802, USA}
\affiliation{Institute for Gravitation and the Cosmos, The Pennsylvania State University, University Park, PA 16802, USA} 

\received{2024 February 2}
\revised{2024 April 11}
\accepted{2024 April 12}

\begin{abstract}

We present the rest-frame ultraviolet-optical spectral properties of 65 broad absorption line (BAL) quasars from the Gemini Near Infrared Spectrograph-Distant Quasar Survey (GNIRS-DQS). These properties are compared with those of 195 non-BAL quasars from GNIRS-DQS in order to identify the drivers for the appearance of BALs in quasar spectra.
In particular, we compare equivalent widths and velocity widths, as well as velocity offsets from systemic redshifts, of principal emission lines. In spite of the differences between their rest-frame ultraviolet spectra, we find that luminous BAL quasars are generally indistinguishable from their non-BAL counterparts in the rest-frame optical band at redshifts $1.55 \lesssim z \lesssim 3.50$. We do not find any correlation between BAL trough properties and the H$\beta$-based supermassive black hole masses and normalized accretion rates in our sample. 
Considering the Sloan Digital Sky Survey quasar sample, which includes the GNIRS-DQS sample, we find that a monochromatic luminosity at rest-frame 2500 \AA\ of \hbox{$\gtrsim 10^{45}$ erg s$^{-1}$} is a necessary condition for launching BAL outflows in quasars.
We compare our findings with other BAL quasar samples and discuss the roles that accretion rate and orientation play in the appearance of BAL troughs in quasar spectra.

\end{abstract}

\keywords{galaxies: active — quasars: emission lines — quasars — broad absorption lines}

\section{Introduction} \label{sec:intro}

Active galaxies are distinguished from quiescent galaxies by the presence of rapidly accreting supermassive black holes (SMBHs) in their centers. The growth of SMBHs over cosmic history appears to be linked to the buildup of their host galaxies, with more massive galaxies generally harboring more massive SMBHs (e.g., \citealt{Magorrian98}, \citealt{Marconi03}, \citealt{DiMatteo05}, \citealt{Yang18}).

Quasars, the most powerful active galactic nuclei (AGNs), can affect their host galaxies by displacing gas and also depositing energy in the form of outflows (e.g., \citealt{Silk98}; \citealt{Cattaneo05}; \citealt{Hopkins05}; \citealt{Begelman06}; \citealt{Hu06}). Observationally, such outflows are manifested by 10-15\% of luminous quasars showing signatures of broad absorption lines (BALs) in their rest-frame ultraviolet (UV) spectra (e.g., \citealt{Reichard03}; \citealt{Trump06}; \citealt{Ganguly07b}; \citealt{Knigge08}; \citealt{Gibson09}). These BAL features are typically associated with high-ionization emission lines (HiBALs; e.g. C~{\sc iv} $\lambda$1549, Si~{\sc iv} $\lambda$1393, N~{\sc v} $\lambda$1240) that are generally blueshifted with respect to systemic redshifts ($z_{\rm sys}$), perhaps partly due to outflows from accretion-disk winds (e.g., \citealt{Murray95}; \citealt{Richards11}). Low-ionization BAL quasars (LoBALs; $\sim$10\% of the BAL-quasar population; e.g., \citealt{Voit93}; \citealt{Trump06}), on the other hand, are characterized by the presence of additional low-ionization species (e.g., Mg~{\sc ii} $\lambda$2803, Al~{\sc iii} $\lambda$1857) in their spectra.

Traditionally, BAL quasars have been defined by the presence of BAL troughs with a minimum velocity width of 2000 km s$^{-1}$ at 10\% depth below the UV continuum \citep[hereafter, W91]{Wey91}. However, it is important to note that W91 defined this index based on low-resolution and low signal-to-noise (S/N) ratio Large Bright Quasar Survey spectra. As a result, the absence of the BAL designation in lower-quality data does not guarantee that a quasar is not a BAL quasar (see Section \ref{sec:sample selection}).

Other common characteristics of BAL quasars include evidence for significant dust reddening  of the continuum, extinction in their UV-optical spectra (e.g., \citealt{Sprayberry92}; \citealt{Richards02}; \citealt{Reichard03}; \citealt{Trump06}), and the vast majority of BAL quasars appear to have low observed X-ray fluxes, with respect to predicted values, based on their optical fluxes (e.g., \citealt{Brandt00}; \citealt{Luo14}). 
While some BAL quasars may be intrinsically X-ray weak, evidence generally suggests that X-ray weakness in these sources is primarily due to obscuration (e.g., \citealt{Gall02}, \citeyear{Gall06}; \citealt{Liu18}; \citealt{Wang22}; but, see also \citealt{Teng14}; \citealt{Morabito14}).

Additionally, the fraction of quasars displaying BALs drops precipitously among the most radio-loud (RL\footnote{We define radio-loud quasars as sources having radio-loudness values of $R > 100$, where $R$ is the ratio of the flux densities at 5 GHz and 4400\AA; \citet{Kellermann89}}) objects (e.g., \citealt{Stocke92}; \citealt{Brotherton98}; \citealt{Becker01}; \citealt{Richards11}). There are indications that this trend may differ for radio-intermediate sources ($10<R<100$) where the fraction of BAL quasars increases with radio luminosity (e.g., \citealt{Petley24}; but, see also \citealt{Calisto23}).
BAL features can also introduce $z_{\rm sys}$ uncertainties; however, masking the locations of these features decreases the redshift errors by about 1\% and the number of catastrophic redshift errors by about 80\% (e.g., \citealt{Garcia23}).

There have been claims that BAL quasars are simply ``normal" quasars observed along preferential lines of sight that penetrate outflowing gas (e.g., W91; \citealt{Ogle91}; \citealt{Schmidt99}; \citealt{Elvis2000}; \citealt{Brotherton06}; \citealt{DiPompeo12}; \citealt[hereafter, R20]{Rankine20}). However, it is also evident that BAL quasars may signify a distinct evolutionary stage, where ``evolution" refers to changes in the spectral energy distribution (SED). During this stage, absorbing material with a high covering fraction is expelled from the central regions of the quasar (e.g., \citealt{Voit93}; \citealt{Becker97}; \citealt{Gregg06}; \citealt{Lipari06}; \citealt{Urrutia09}). R20 further emphasizes that the structure of BALs varies across the observed parameter space, and different SEDs lead to different mean BAL properties. Observationally, careful comparisons of the rest-frame UV emission-line properties reveal that BAL and non-BAL quasars appear to be drawn from the same parent population of quasars (e.g., W91; \citealt{Reichard03}; \citealt{Baskin13}). In addition, \citet{Gall07} highlights that the mid-infrared properties of BAL quasars are also consistent with those of non-BAL quasars of comparable luminosity.

While in the early studies, models were often employed to suggest either orientation or evolution as explanations for BAL properties, it becomes clear that a combination of evolutionary and orientation effects, along with other factors, must contribute to the observed phenomena. Consequently, investigating the underlying physics becomes crucial for a comprehensive understanding of the structure, evolution, and feedback mechanisms among all quasars and their host galaxies.

This paper aims to analyze rest-frame UV-optical spectra of BAL and non-BAL quasars, comparing emission-line properties and velocity offsets from $z_{\rm sys}$ to shed light on the drivers for the appearance of BALs in quasar spectra. Prior studies, such as those by W91 and \citet{Reichard03}, have delved into similar investigations, examining rest-frame UV spectra between BAL and non-BAL quasars. Our study focuses primarily on comparing rest-frame optical properties between the two groups of quasars. A statistically meaningful comparison of this kind is now possible by utilizing the Gemini Near Infrared Spectrograph–Distant Quasar Survey (GNIRS-DQS) which includes 65 (195) luminous BAL (non-BAL) quasars (\citealt{M21}, \citeyear{M23}, hereafter, M21, M23, respectively).

The structure of this paper is as follows. In Section \ref{sec:sample selection}, the sample selection, and methodology are described. Section \ref{sec:results} presents our main results. Section \ref{sec:discussion} discusses our main findings with comparisons to other work. Section \ref{sec:conclusion} presents our summary and conclusions. Throughout this work, we adopt a $\Lambda$CDM cosmology with $H_{0}=70$ km s$^{-1}$ Mpc$^{-1}$, $\Omega_{M}=0.3$, and $\Omega_{\Lambda}=0.7$ when calculating quantities such as quasar luminosities or luminosity distances (e.g., \citealt{Spergel07}).

\section{Sample Selection} \label{sec:sample selection}

A pre-requisite for any investigation of the statistical distribution of emission-line properties is a large, well-defined source catalog. To investigate differences in the rest-frame optical properties of quasars with BAL features compared to those without, one generally must resort to sources above $z\sim1.5$. Below that threshold, the scarcity of rest-frame UV spectra poses a challenge as these spectra are primarily available through the Hubble Space Telescope (HST) archive in a non-uniform fashion and based on numerous selection criteria. Consequently, a more cost-effective approach is to conduct the study at higher redshifts where uniform catalogs of rest-frame UV spectra are more abundant, and rest-frame optical spectra can be obtained more economically from near-infrared (NIR) spectroscopy. 

For this reason, we selected sources from the GNIRS-DQS catalog which constitutes the largest, uniform inventory of rest-frame optical spectral properties of luminous quasars at high redshift from the Sloan Digital Sky Survey (SDSS, \citealt{York00}). Specifically, the catalog consists of 260 quasars, 65 of which are BAL quasars, having $-28.0 \lesssim M_i \lesssim -30.0$ mag with redshifts between $1.55 \lesssim z \lesssim 3.50$ (to center the H$\beta$+[O~{\sc iii}] spectral complex in the $J$, $H$, and $K$ bands) with monochromatic luminosities ($\lambda L_{\lambda}$) at rest-frame 5100 \AA\ (hereafter, $\lambda L_{5100}$) in the range of $\sim 10^{46} - 10^{47}$ erg~s$^{-1}$. Our sample consists primarily of HiBAL quasars with four LoBAL quasars previously identifed by \citet{Trump06} or confirmed through visual inspection. The small number of LoBAL quasars prevents us from conducting a statistically meaningful comparison with their HiBAL counterparts.

The GNIRS-DQS selection criteria may have excluded sources known to exhibit some degree of obscuration. Thus, there is a possibility that a certain number of luminous SDSS BAL quasars did not meet the brightness threshold of GNIRS-DQS. Additionally, there is also a possibility that the GNIRS-DQS sample is biased toward more face-on systems given the high luminosities of these sources (e.g., \citealt{Runnoe13}; \citealt{DiPompeo14}). Yet, in such a scenario, we would have anticipated a larger fraction of RL quasars and, correspondingly a lower fraction of BAL quasars. 

As mentioned in Section \ref{sec:intro}, RL quasars are a small fraction of the overall quasar population, and this fraction is even lower among BAL quasars. While 17 non-BAL quasars in our sample are formally RL, only one of the BAL quasars, SDSS J114705.24+083900.6, is also a RL quasar; in total, RL quasars consitute $\sim$ 7\% of the entire GNIRS-DQS sample. If our sample is indeed biased toward more face-on systems, correcting for this orientation bias would likely result in a higher BAL quasar fraction than we measure ($65/260$ = 25\%), which is a less probable scenario. Furthermore, we also examine whether our sample is biased with respect to radio-loudness. This investigation involves analyzing photometric data from the Wide-field Infrared Survey Explorer (WISE; e.g., \citealt{Wright10}), specifically focusing on W1 and W2 magnitudes with respect to radio-loudness. We find no discernible differences between BAL and non-BAL quasars, or between RL and non-RL quasars, across the NIR continuum of these sources. Therefore, we do not have evidence that our sample is biased with respect to quasar orientation or radio-loudness.

The spectral measurements in this work were obtained from the catalog of spectroscopic properties of the GNIRS-DQS quasars in M23. M23 determined the $z_{\rm sys}$ values for GNIRS-DQS sources based on the available emission line that carried the smallest intrinsic uncertainty (see, \citealt{Boroson05}; \citealt{Shen16}; \citealt{Nguyen20}). In order of increasing uncertainty, these values are obtained from [O~{\sc iii}] $\lambda$5007, Mg~{\sc ii} $\lambda$$\lambda$2798, 2803, and H$\beta$ $\lambda4861$. For those objects where [O~{\sc iii}] is too weak to obtain a reliable $z_{\rm sys}$ measurement, Mg~{\sc ii} and H$\beta$ are substituted as suitable proxies (see, M23).

The SMBH masses ($M_{\rm BH}$) and mass-weighted accretion rates (i.e., Eddington ratios, or $L/L_{\rm Edd}$ values) for all sources in this work were obtained from \citet{Dix23} and \citet[hereafter, D23 and H23, respectively]{Ha23} following the prescription in \citet{Maithil22}. Briefly, these values are single-epoch $M_{\rm BH}$ estimates using the H$\beta$ line coupled with corrections based on the Fe~{\sc ii} emission strength in the rest-frame wavelength range $\lambda$$\lambda$4434-4684 (see, \citealt{Du19}).

Rest-frame optical properties for the 65 BAL quasars taken from M23 are reported in Table \ref{tab:table1}. Column (1) provides the source name; Column (2) gives the redshift based on H$\beta$ ($z_{\rm H\beta}$); Column (3) gives the redshift based on [O~{\sc iii}] ($z_{[\rm O~{\rm III}]}$); Column (4) gives the velocity offset between the two redshifts; Column (5) gives rest-frame equivalent width (EW) of the H$\beta$ line; Column (6) gives rest-frame EW (Fe~{\sc ii}); Column (7) gives full-width at half maximum (FWHM) intensity of H$\beta$; Column (8) gives rest-frame EW ([O~{\sc iii}]); Column (9) gives FWHM ([O~{\sc iii}]); Column (10) gives asymmetry of [O~{\sc iii}] $\lambda 5007$; Column (11) gives log $\lambda L_{5100}$; Column (12) gives Fe~{\sc ii}-corrected H$\beta$-based $M_{\rm BH}$ estimates taken from D23; Column (13) gives Fe~{\sc ii}-corrected H$\beta$-based $L/L_{\rm Edd}$ values taken from H23.

Rest-frame UV properties for the 65 BAL quasars are reported in Table \ref{tab:table2}. Column (1) provides the source name; Column (2) gives the visually inspected redshift, $z_{\rm vi}$, from SDSS Data Release 16 (DR16; \citealt[Table D1, Column 17]{Lyke20}); Column (3) gives the systemic redshift from M23; Column (4) gives the velocity offset between the two redshifts; Column (5) gives the C~{\sc iv} emission-line rest-frame EW from R20 for 52 BAL sources; Column (6) gives C~{\sc iv} Blueshifts from R20. Additionally, Column (7) gives the Balnicity Index (BI) defined by W91 for C~{\sc iv}, as: 

\begin{equation}
    {\rm BI} = \int_{3000}^{25000} \left( 1 - \frac{f(v)}{0.9}\right)~{C}~{\rm d} v ~\rm{(km~s^{-1})}
\end{equation}

\hfill

\noindent where $f(v)$ is the continuum-normalized flux as a function of velocity, $v$, relative to the line center. The constant $C$ equals 1 in regions where $f(v) < 0.9$ for at least \hbox{${2000~\rm{km~s^{-1}}}$}, counting from large to small outflow velocities, otherwise $C = 0$. The outflow velocity, $v$, is defined to be negative for blueshifts. BI is the traditionally used metric to separate BAL quasars from their non-BAL counterparts. 
In addition to BI, Column (8) gives the Absorption Index (AI) defined by \citet{Hall02}, as: 

\begin{equation}
    {\rm AI} = \int_{0}^{25000} \left( 1 - \frac{f(v)}{0.9}\right)~{C}~{\rm d} v ~\rm{(km~s^{-1})}
\end{equation}

\hfill

\noindent to account for uncertainties in the systemic redshift, the continuum shape, and to measure intrinsic absorption systems. Here, $C$ equals 1 in regions where $f(v) < 0.9$  continuously for at least \hbox{${450~\rm{km~s^{-1}}}$}. The BI and AI values in Table \ref{tab:table2} were obtained from the literature (Column 9); eleven sources that have no measurable  BI values with \hbox{AI $> 0$} are marked with \hbox{BI = 0}, two sources have both \hbox{BI = 0} and \hbox{AI = 0}, and one source with no measurable AI value with \hbox{BI $> 0$} is marked with \hbox{AI = 0} (see Table \ref{tab:table2}). All these sources have been visually classified as BAL quasars (see M23). Table \ref{tab:table3} provides the mean ($\mu$), standard error on the mean (SEM), standard deviation ($\sigma$), and median (Med) with bootstrapped uncertainty estimates for the properties listed in Table \ref{tab:table1}, as well as associated probability ($p$-value) of statistical tests (see below) for BAL and non-BAL sources in columns \hbox{2 - 12}, respectively.

\section{Analysis and Results} \label{sec:results}

\subsection{Rest-frame optical properties}\label{subsec:optical}

\begin{figure*}[ht!]
\centering
\includegraphics[width=17cm]{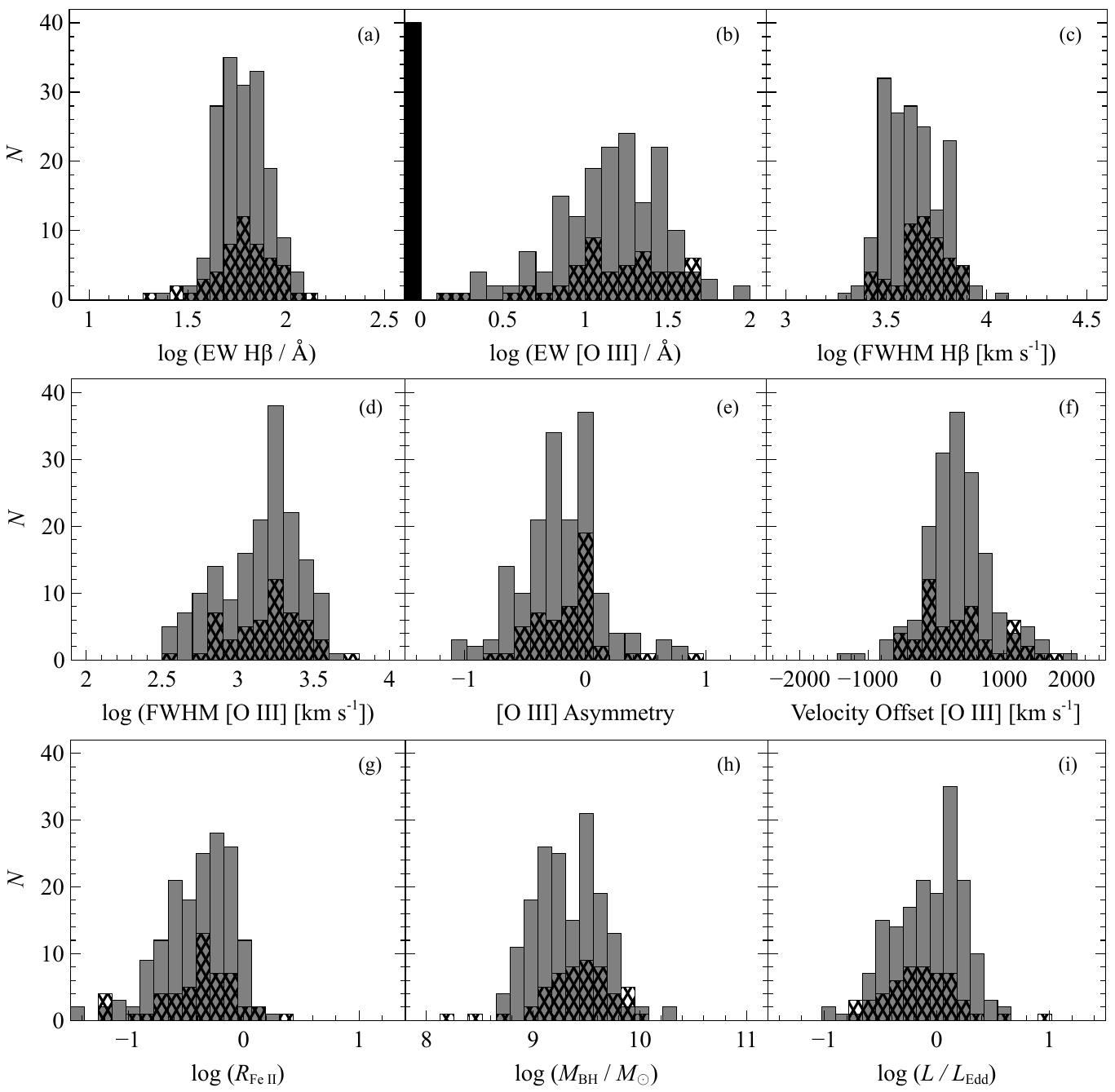}
\caption{Distributions of the optical emission-line properties between BAL (hatched) and non-BAL (grey) quasars from the GNIRS-DQS sample. The comparison of all distributions reveals that the two populations are largely similar (see Section \ref{sec:results} and Table \ref{tab:table3} for statistical analysis). For forty sources (13 BAL and 27 non-BAL quasars) that did not meet the M23 threshold of reliability for an EW ([O~{\sc iii}]) measurement (i.e., greater than 1 \AA), we place upper limits of 1 \AA\ on their EW values (solid black column in panel b; see also Table \ref{tab:table1}). Top row (panels a, b, and c) includes distributions of EW (H$\beta$), EW ([O~{\sc iii}]), and FWHM (H$\beta$) of 65 BAL and 195 non-BAL quasars. Middle row (panels d, e, and f) includes [O~{\sc iii}] FWHM, asymmetry and velocity offset distributions for only 52 BAL and 168 non-BAL quasars that meet the M23 EW ([O~{\sc iii}]) threshold. Bottom row (panels g, h, i) includes distributions of $R_{\rm Fe~{\rm II}}$, $M_{\rm BH}$ and $L/L_{\rm Edd}$. 
\label{fig:Figure 1}}
\end{figure*}

\begin{figure}[ht!]
\centering
\includegraphics[width=8 cm]{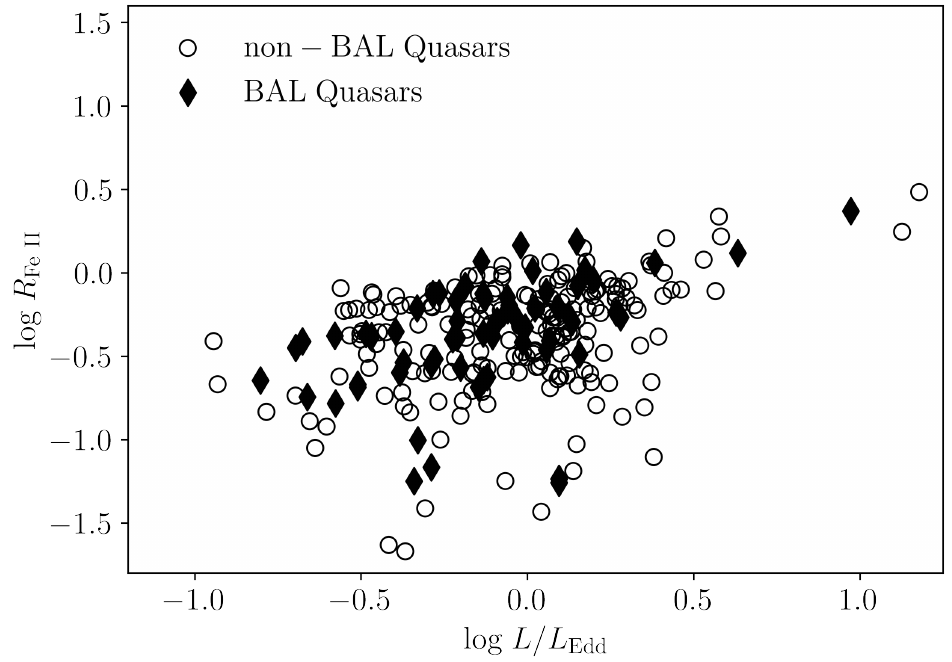}
\caption{$R_{\rm Fe~{\rm II}}$ versus $L/L_{\rm Edd}$ for non-BAL (open circles) quasars and BAL (diamonds) quasars from GNIRS-DQS. Both groups reveal a strong positive correlation between the two parameters (see Section \ref{subsec:optical}).
\label{fig:Figure 2}}
\end{figure}

We compare basic properties for multiple rest-frame optical emission lines between the 65 BAL and 195 non-BAL sources in GNIRS-DQS in Figure \ref{fig:Figure 1}. This figure includes a comparison of the rest-frame EWs of the H$\beta$ and [O~{\sc iii}] emission lines. The means and medians are consistent within their errors for both distributions (see Table \ref{tab:table3}). It is evident that at $1.55 \lesssim z \lesssim 3.50$ a large fraction of BAL quasars have [O~{\sc iii}] emission which is just as strong as for the non-BAL quasars in the GNIRS-DQS sample. This result is likely due to a narrow luminosity range inherent in the GNIRS-DQS sample.

By design, GNIRS-DQS targeted highly luminous quasars, biased toward having higher $L/L_{\rm Edd}$ values. The [O~{\sc iii}] emission in luminous high-redshift quasars is relatively weak (whether BAL or not, due to the Baldwin Effect; \citealt{Baldwin77}) and, similarly, Fe~{\sc ii} emission is relatively strong in such sources (e.g., \citealt{Dietrich02}; \citealt{Netzer04}; \citealt{Shen16}; M21; M23). At lower redshifts, however, a larger range of parameter space is observed (e.g., a broader range of quasar luminosities).

To check whether the BAL outflows are manifested in the sources' [O~{\sc iii}] emission line properties, we compare the FWHM and asymmetry
values of the [O~{\sc iii}] $\lambda5007$ emission line (taken from Table \ref{tab:table1}) between BAL and non-BAL quasars in Figure \ref{fig:Figure 1}. The mean and median FWHM ([O~{\sc iii}]), as well as the asymmetry values for BAL and non-BAL quasars are consistent within their errors (see Table \ref{tab:table3}).
We also computed the velocity offset ($\Delta v$) between the $z_{\rm H\beta}$ and $z_{[\rm O~{\rm III}]}$ values of a source between BAL and non-BAL quasars, using

\begin{equation}
    \Delta v = c~\left(\frac{z_{\rm H\beta} - z_{[\rm O~{\rm III}]}}{1 + z_{[\rm O~{\rm III}]}}\right)
\end{equation}

\hfill

\noindent where $c$ is the speed of light in km s$^{-1}$. The distributions of $\Delta v$ values for BAL and non-BAL quasars are presented in Figure \ref{fig:Figure 1}.

The relative intensity of the Fe~{\sc II} emission blend with respect to the H$\beta$ line appears to be correlated with quasar accretion rate in terms of $L/L_{\rm Edd}$ (e.g., \citealt{Boroson92}; \citealt{Netzer07}, \citealt{Shen&Ho14}, \citealt{Du19}). Using the EWs of the  Fe~{\sc ii} blend and H$\beta$ line of all GNIRS-DQS quasars, a comparison is made between $R_{\rm Fe~{II}}$, defined as EW(Fe~{\sc ii})/EW(H$\beta$), of BAL and non-BAL quasars as shown in Figure \ref{fig:Figure 1}. The mean and median $R_{\rm Fe~{II}}$ values for BAL and non-BAL quasars are consistent within their errors (see Table \ref{tab:table3}). This result indicates that BAL and non-BAL quasars, in this sample of highly luminous sources, have similar $L/L_{\rm Edd}$ distributions despite the presence of outflows in the former.

A similar comparison was performed between the FWHM values of the H$\beta$ line between the BAL and non-BAL quasars and is displayed in Figure \ref{fig:Figure 1}. The figure also compares the $M_{\rm BH}$ values, which show a similar distribution to the FWHM (H$\beta$) distribution, as the $M_{\rm BH}$ values depend primarily on FWHM (H$\beta$) for sources in a relatively narrow luminosity range (e.g., D23). The comparison of the Eddington ratios ($L/L_{\rm Edd}$, where $L$ is the bolometric luminosity), obtained from M23 and H23, for our populations of BAL and non-BAL quasars, presented in Figure \ref{fig:Figure 1}, shows no significant differences between the two groups of sources. Figure \ref{fig:Figure 2} shows a correlation between $R_{\rm Fe~{II}}$ and $L/L_{\rm Edd}$ values for the GNIRS-DQS BAL and non-BAL quasars. We find a Spearman rank correlation coefficient of 0.54 ($p = 3.48 \times 10^{-6}$) and 0.35 ($p = 4.58 \times 10^{-7}$) for BAL and non-BAL quasars, respectively. Both groups exhibit a statistically significant positive correlation between $R_{\rm Fe~{II}}$ and $L/L_{\rm Edd}$.

To test if the rest-frame optical properties of BAL quasars are significantly different from their non-BAL counterparts, we ran a two-tailed Kolmogorov-Smirnov (K-S) test on all the parameter distributions shown in Figure \ref{fig:Figure 1}. We choose a cutoff for ‘significant’ differences in distributions of $p$ = 0.05 and $p$ = 0.01 to indicate rejection or failure of rejection of the null hypothesis at 95\% and 99\% significance level, respectively. Test results with $p$-values lower than these limits are possibly indicative of different parent populations for BAL and non-BAL quasars, and are highlighted in bold in Table \ref{tab:table3}.

To run the K-S test on the distributions of EW ([O~{\sc iii}]), we removed 13 BAL quasars and 27 non-BAL quasars that did not meet the M23 threshold of reliability for EW([O~{\sc iii}]) values greater than 1 \AA. In each test the hypothesis that properties from both samples, BAL and non-BAL quasars, arise from the same parent population could not be rejected at a significance level of 95\% or 99\% (see Table \ref{tab:table3}).

The K-S test may not effectively capture discrepancies in the tails of distributions. To address this issue, we also conducted a two-tailed Anderson-Darling (A-D) test on all the parameter distributions depicted in Figure \ref{fig:Figure 1}, utilizing the same significance levels as employed in the K-S test. The A-D test exhibits greater sensitivity to differences in the tails of distributions and is capable of detecting even minute distinctions, particularly in large sample sizes. Our findings indicate that the null hypothesis, that both BAL and non-BAL quasars arise from the same parent population of sources, is rejected for FWHM (H$\beta$) and $M_{\rm BH}$ at both the 95\% and 99\% significance levels, and for $L/L_{\rm Edd}$ at only the 95\% significance level. However, the null hypothesis was not rejected at either significance level for EW (H$\beta$), EW ([O~{\sc iii}]), FWHM ([O~{\sc iii}]), asymmetry ([O~{\sc iii}]), velocity offset ([O~{\sc iii}]), or $R_{\rm Fe~{II}}$.
In particular, we do not detect any manifestation of BAL outflows in the [O~{\sc iii}] line properties of our sources.

We conducted a Monte Carlo simulation on properties where the null hypothesis was rejected with the A-D test, specifically focusing on $L/L_{\rm Edd}$, $M_{\rm BH}$, and FWHM (H$\beta$). In this simulation, we performed 10,000 realizations to account for uncertainties associated with each value. For approximately 90\% of the 10,000 realizations, the A-D test was unable to reject the null hypothesis in each test. The simulation indicates that the A-D test is particularly sensitive to the tails of the distributions rather than the core values of $L/L_{\rm Edd}$, $M_{\rm BH}$, and FWHM (H$\beta$). Therefore, since each value carries a degree of uncertainty, it is important to interpret the A-D test results with caution. 

If indeed $M_{\rm BH}$, FWHM (H$\beta$), and possibly $L/L_{\rm Edd}$ values of BAL and non-BAL quasars were found to originate from distinct parent populations, then this would have implied that the underlying mechanisms governing the broad-line region (BLR) dynamics and central SMBH accretion processes differ between these two groups of quasars. Any putative differences may have required refining mass estimation prescriptions to include additional BLR dynamics such as outflows. However, given that the differences are marginal at best, these properties remain broadly indistinguishable between BAL and non-BAL quasars.

\begin{figure*}[ht!]
\centering
\includegraphics[width=7 cm]{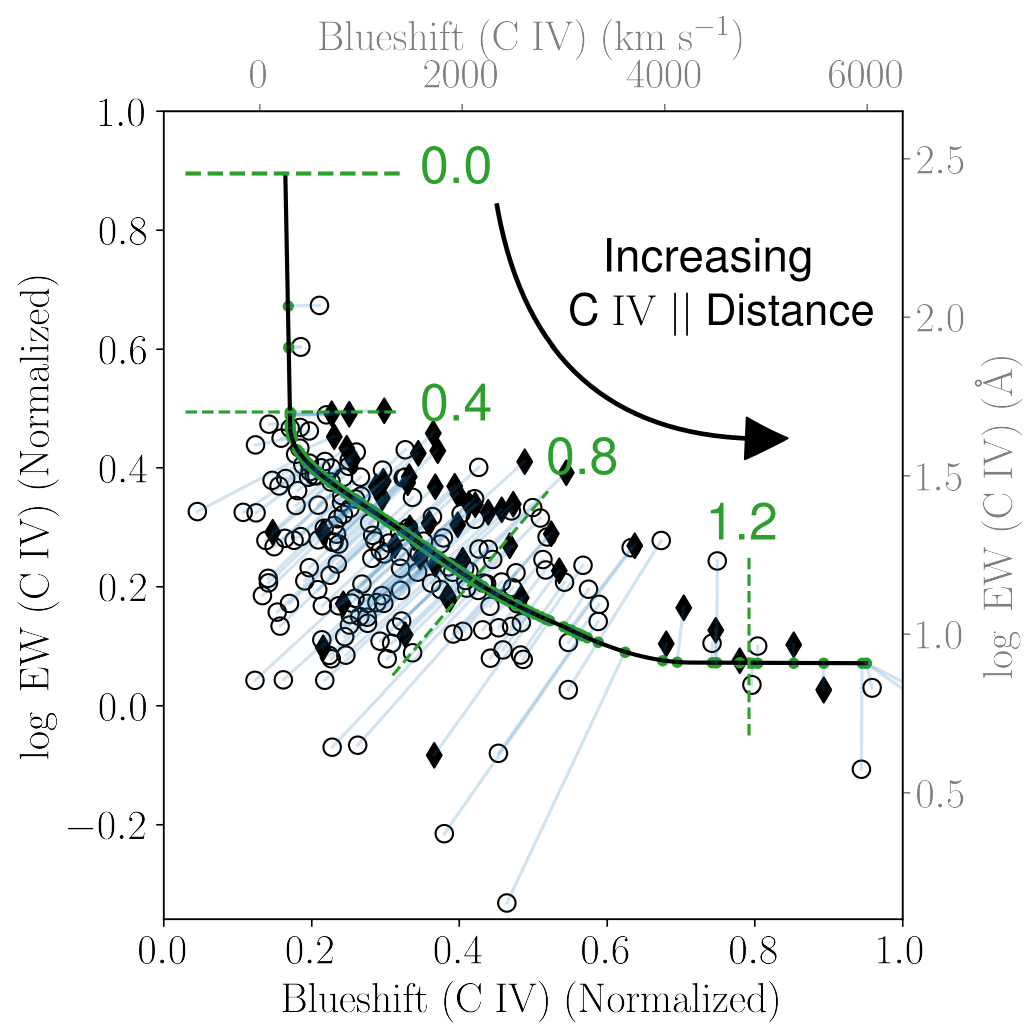}
\includegraphics[width=8.4 cm]{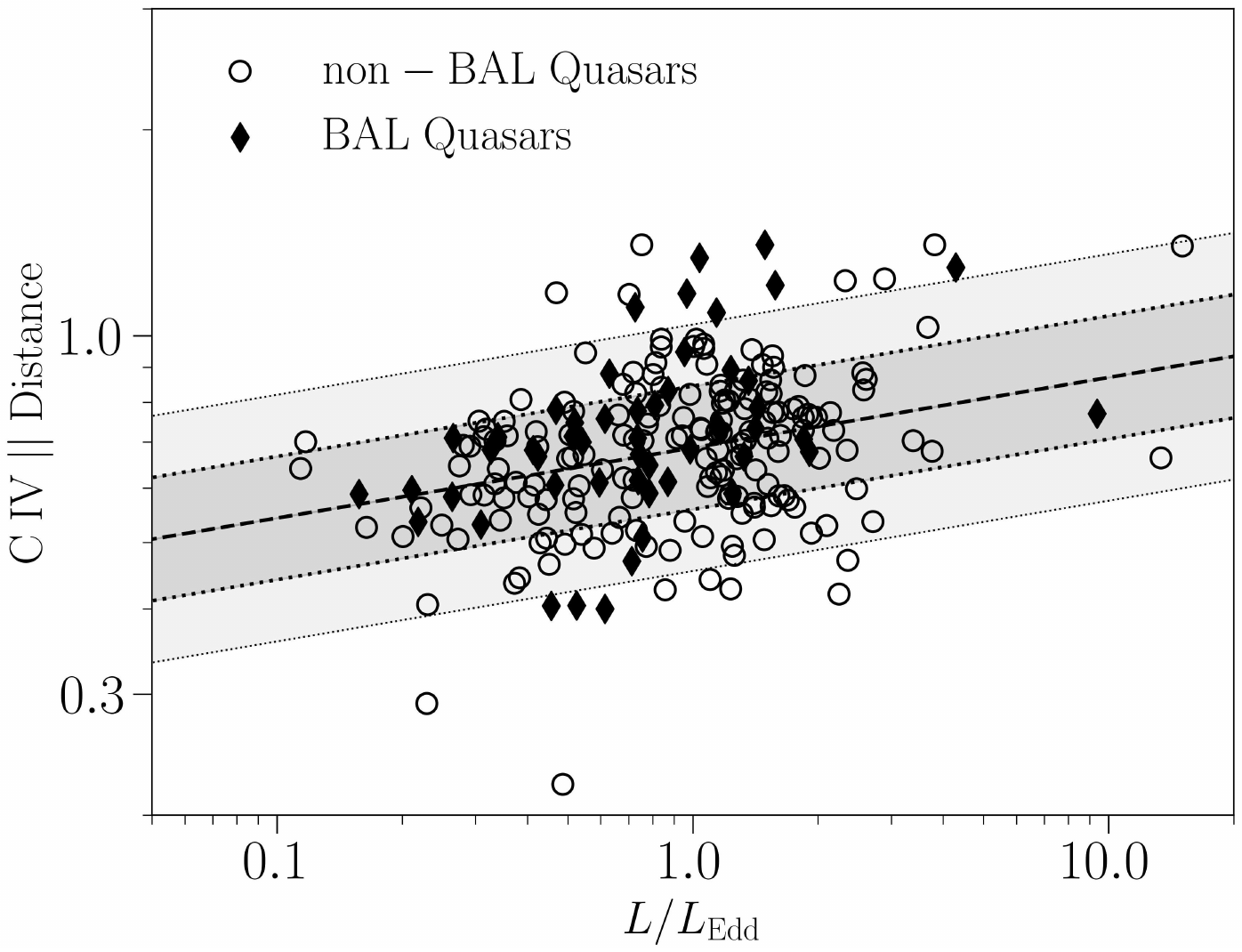}
\caption{Left panel: distribution of EW (C~{\sc iv}) vs. blueshift (C~{\sc iv}) for 177 non-BAL quasars and 52 BAL quasars; symbols are the same as in Figure \ref{fig:Figure 2}. The data are first normalized so that the two axes share the same limit; each data point is then projected (thin blue lines) onto the best-fit curve obtained from \citet{Rivera22}. The \hbox{C~{\sc iv} $||$ Distance} value of each quasar is defined as its projected position (green point) along the solid black curve. Representative \hbox{C~{\sc iv} $||$ Distance} values are indicated by dashed lines and numbers. The ranges of C~{\sc iv} blueshift and EW are $\rm -1000~to~6000~km~s^{-1}$ and $\rm 2~to~140$ \AA, respectively. Right panel: \hbox{C~{\sc iv} $||$ Distance} parameter versus Fe~{\sc ii}-corrected H$\beta$-based $L/L_{\rm Edd}$. The correlation for the non-BAL quasar sample, obtained by fitting a linear model, is indicated by a dashed line. The dark (light) shaded region represents the 1$\sigma$ (2$\sigma$) deviation from the fitted correlation (see Section \ref{sec:results}).
\label{fig:Figure 3}}
\end{figure*}

\subsection{Rest-frame UV properties} \label{subsec:UV}

To explore the C~{\sc iv} parameter space of BAL and non-BAL quasars, we compare a sample of 177 GNIRS-DQS non-BAL quasars with a sample of 52 BAL quasars with reliable C~{\sc iv} measurements taken from D23 and R20, respectively. D23 provides measurements of the C~{\sc iv} emission-line properties of 177 GNIRS-DQS sources, following the exclusion of 64 BAL quasars, 16 RL quasars, one BAL and RL quasar, and two sources that do not have reliable C~{\sc iv} coverage. For the BAL quasars, we adopted C~{\sc iv} measurements (EW and blueshift) from R20 for 52 of our sources (not including the single RL BAL quasar, SDSS J114705.24+083900.6) that appeared in their sample. 

For a detailed description of the C~{\sc iv} measurements in our BAL quasars, we refer the reader to Section 6.1 of R20. In summary, calculation of the C~{\sc iv} emission line parameters followed a non-parametric approach. Initially, a power-law continuum was subtracted from a reconstructed spectrum, after which the emission line flux was integrated to determine the EW, and the blueshift was calculated from the wavelength that bisects the total cumulative line flux relative to the systemic redshift of the quasar. Wavelength regions of the spectra affected by BAL troughs were masked, covering areas blueward of the C~{\sc iv} emission line (i.e., 1430-1546 \AA), resulting in a relatively symmetric emission line profile.

The left panel of Figure \ref{fig:Figure 3} displays normalized values of the C~{\sc iv} emission line EW versus blueshift for the 177 non-BAL quasars compared to 52 BAL quasars from GNIRS-DQS. While these two parameters, on their own, are not ideal accretion-rate indicators (e.g., \citealt{Richards11}; \citealt{Shemmer15}), a combination of these two parameters appears to provide a robust indication of the accretion rate for all quasars as manifested in the right panel of Figure \ref{fig:Figure 3} (e.g., H23). This consolidated {C~{\sc iv} parameterization, termed the \hbox{C~{\sc iv} $||$ Distance}, indicates the projected location onto a nonlinear first principal component of the EW-blueshift diagram and the piecewise polynomial best-fit curve traces the C~{\sc iv} parameter space of sources across wide ranges of redshifts and luminosities. To calculate the \hbox{C~{\sc iv} $||$ Distance} parameter, we follow the procedure summarized in \citet{Rivera22} and detailed in \citet{Mcaffrey21}.

Upon inspection of Figure \ref{fig:Figure 3}, it appears as if, while BAL and non-BAL quasars demonstrate broad similarities within the C~{\sc iv} parameter space, the former tend to concentrate toward the top right above the best-fit curve in the left panel. We quantitatively tested this observation by also comparing the \hbox{C~{\sc iv} $\perp$ Distance} parameter (the perpendicular to the \hbox{C~{\sc iv} $||$ Distance} parameter), also defined in \citet{Rivera22}, between BAL and non-BAL quasars using a K-S test. The \hbox{C~{\sc iv} $\perp$ Distance} parameter was measured relative to the closest (projected) point on the best-fit curve. The results indicate the null hypothesis, that BAL and non-BAL quasars arise from the same parent population, can be rejected at significance levels of 95\% and 99\%. We suspect this distinction between BAL and non-BAL quasars stems from the fact that measurements for the BAL quasars were performed by R20, while those for the non-BAL quasars were performed by D23, relying on $z_{\rm sys}$ values from M23.

In the case of \hbox{C~{\sc iv} $||$ Distance} versus $L/L_{\rm Edd}$, we find a Spearman rank correlation coefficient of 0.52 ($p = 7.65 \times 10^{-5}$) and 0.28 ($p = 1.47 \times 10^{-4}$) for BAL and non-BAL quasars, respectively. This result indicates that despite being a smaller subset of sources with potentially large uncertainties in the C~{\sc iv} measurements, the BAL quasars exhibit a stronger H$\beta$-based $L/L_{\rm Edd}$ and \hbox{C~{\sc iv} $||$ Distance} correlation compared to non-BAL quasars. We fit a linear model to the \hbox{C~{\sc iv} $||$ Distance} parameter and $L/L_{\rm Edd}$ space, taking into account only the non-BAL quasars. The mean deviation of the BAL quasars from the best-fit line is $\sim 1.05\sigma$.

We also searched for any potential relationship of BI and AI as a function of EW ([O~{\sc iii}]) and $L/L_{\rm Edd}$ values. However, we could not identify any significant correlation or trend amongst these parameters. We further performed an investigation of trough properties, including maximum and minimum velocity of each trough ($V_{\rm max}$, $V_{\rm min}$), and width of each trough ($V_{\rm width}$), as a function of H$\beta$-based $L/L_{\rm Edd}$ and $M_{\rm BH}$. The trough values were taken from R20 (see Table 1 therein) for 52 of our BAL sources that were part of their significantly larger sample. Our analysis does not reveal any significant correlation or discernible trend between the BAL trough properties and $L/L_{\rm Edd}$ or $M_{\rm BH}$. This lack of clear trend could be attributed to the constraints imposed by a small sample size and limitations inherent in deriving comprehensive information from basic measurements such as those obtained from R20 (see also, \citealt{Leighly22}).

To test whether BAL quasars exhibit significantly different blueshifts compared to non-BAL quasars, we compute the $\Delta v$ values between the $z_{\rm vi}$ value from \citet{Lyke20}\footnote{We note that for~20\% of our BAL quasars, the SDSS Pipeline fails to provide a reliable redshift (\citealt{Lyke20}, Table D1, Column 29), thus the $z_{\rm vi}$ values should be adopted. \citet{Lyke20} only provides $z_{\rm vi}$ values for 52 BAL and 150 non-BAL GNIRS-DQS quasars (note that these 52 sources do not entirely coincide with the 52 sources that have C~{\sc iv} measurements in R20).} and the $z_{\rm sys}$ value of a source taken from M23. If there is a noticeable difference, it could suggest the presence of an additional factor that is responsible for driving outflows in BAL quasars. The velocity offset is computed using

\begin{equation}
    \Delta v = c~\left(\frac{z_{\rm vi} - z_{\rm sys}}{1 + z_{\rm sys}}\right)
\end{equation}

\hfill

\noindent in a similar manner to Equation 3 in Section \ref{subsec:optical}. The distributions of $\Delta v$ values for 52 BAL and 150 non-BAL quasars are presented in Figure \ref{fig:Figure 4}; the $\Delta v$ values for the former\footnote{The $z_{\rm vi}$ values of five of the sources are identical to their $z_{\rm sys}$ values. We, therefore, assign an upper limit of $|90|~\rm km~s^{-1}$ for their $\Delta v$ values in Table \ref{tab:table2}; these values are treated as zeroes in Figure \ref{fig:Figure 4} and our analysis.} are taken from Table \ref{tab:table2} and the $\Delta v$ values for the latter are computed using $z_{\rm vi}$ values from \citet{Lyke20} and $z_{\rm sys}$ values from M23. 

The comparison of both distributions of BAL and non-BAL quasars reveals no significant difference between these two populations. We confirm this result by running K-S and A-D tests on the distributions in a similar manner to the tests run in Section \ref{subsec:optical}. Our analysis reveals that the hypothesis asserting that $\Delta v$ from both samples, BAL and non-BAL quasars, come from the same parent population was not rejected at the 95\% or 99\% significance level with the K-S and A-D tests (see Table \ref{tab:table3}). This result highlights the fact that both populations exhibit a high degree of similarity.

\begin{figure}
\centering
\includegraphics[width=7.5cm]{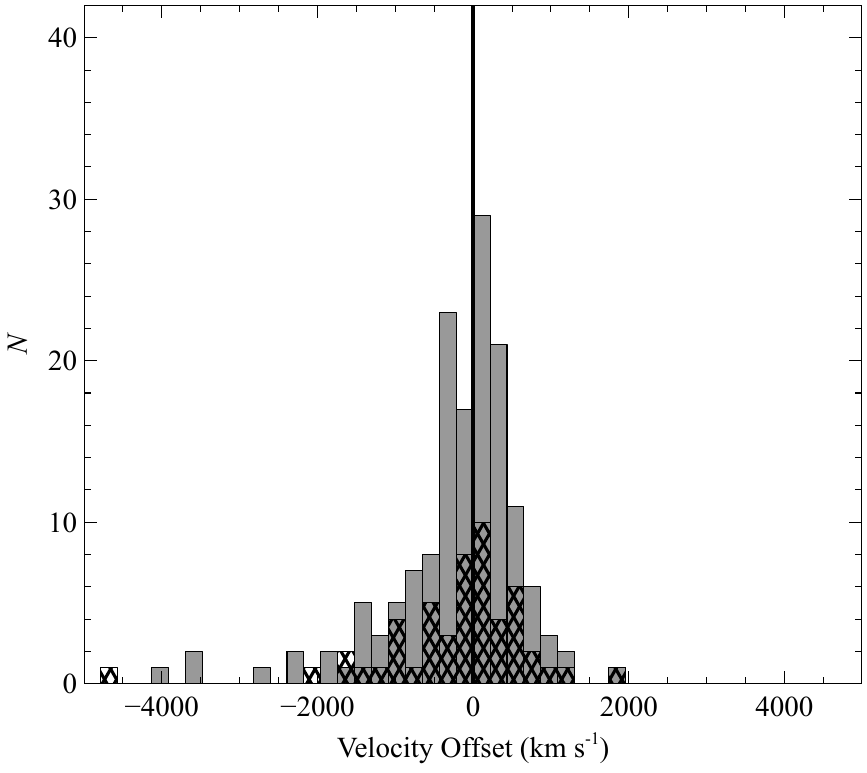}
\caption{Distribution of the velocity offset between $z_{\rm vi}$ from \citet{Lyke20}, and the $z_{\rm sys}$ value of a source between BAL (hatched) and non-BAL (grey) quasars from the GNIRS-DQS sample. Zero velocity offset is indicated by the solid line. Based on statistical tests, both distributions are broadly similar (see Section \ref{sec:results}). 
\label{fig:Figure 4}}
\end{figure}

\raggedbottom
\section{Discussion} \label{sec:discussion}

We find that BAL quasars are generally indistinguishable from their non-BAL counterparts in the rest-frame optical band, 
an extension of the W91 result concerning rest-frame UV emission-line properties.
This conclusion resonates with the results of \citet{Schulze17} who find no discernible differences in the rest-frame optical properties between LoBAL and non-BAL quasars, particularly, with respect to their $M_{\rm BH}$ and $L/L_{\rm Edd}$ values.
To understand the underlying causes behind the appearance of BAL troughs in quasar spectra, we consider the potential influence of accretion rate and orientation.

Our highly uniform sample consists of quasars selected at high redshifts with high luminosities, likely above the luminosity threshold required to launch accretion-disk winds (e.g., \citealt{Laor02}, \citealt{Bieri17}; \citealt{Quera23}). \citet{Giustini19} further argue that the strongest winds are expected to arise in objects with both the highest $M_{\rm BH}$ values and highest accretion rates such as those in the GNIRS-DQS sample (see also, D23; \citealt{Temple23}). This may explain the relatively high fraction of BAL quasars observed in GNIRS-DQS.

\begin{figure}
\centering
\includegraphics[width=8.5cm]{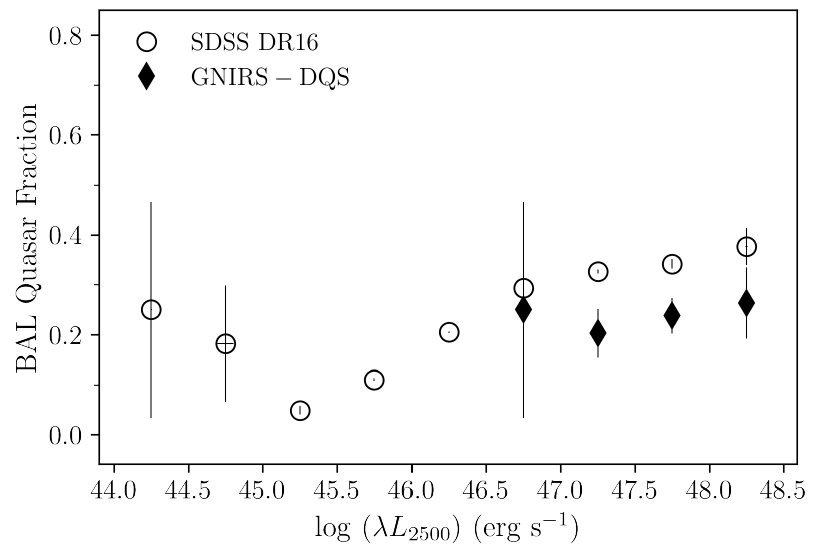}
\caption{Fraction of BAL quasars as a function of $\lambda L_{\lambda}$ at 2500 \AA. Each point represents the average BAL quasar fraction within equal luminosity bins ($\Delta{\lambda L_{\lambda}}$) of 0.5 dex. The SDSS DR16 sample (\citealt{Lyke20}) includes 141,241 quasars between $1.57 \le z \lesssim 3.5$. 
\label{fig:Figure 5}}
\end{figure}

Figure \ref{fig:Figure 5} depicts how the observed BAL quasar fraction in GNIRS-DQS changes as a function of luminosity. For comparison, Figure \ref{fig:Figure 5} also displays similar data for SDSS DR16 quasars, which is the parent sample of the GNIRS-DQS sources. For this comparison sample, we only included sources identified in \citet{Lyke20} as having a BAL probability $\ge 0.75$. We further limited that sample to sources lying in the following redshift ranges, \hbox{$1.57 \le z \lesssim 1.65$}, \hbox{$2.10 \lesssim z \lesssim 2.40$}, and \hbox{$3.20 \lesssim z \lesssim 3.50$}, similar to the GNIRS-DQS redshift intervals; no brightness restriction was applied. Since SDSS DR16 only identifies BAL quasars at \hbox{$z \ge 1.57$}, two GNIRS-DQS sources are excluded from Figure \ref{fig:Figure 5}, $\rm SDSS~J080636.81+345048.5$ at $z = 1.553$ and $\rm SDSS~J090247.57+304120.7$ at $z = 1.560$. For all sources in Figure \ref{fig:Figure 5}, monochromatic luminosities at a rest-frame wavelength of 2500 \AA\ were derived from the PSF $m_{\rm i}$ values in \citet{Lyke20}, assuming a quasar continuum of the form $f_{\nu}\propto \nu^{-0.5}$ (e.g., \citealt{vanden01}). These luminosities are represented by equal bins of ($\Delta{\lambda L_{\lambda}}$) $=$ 0.5 dex, including $\sim 4-60,000$ sources, per bin, with error bars representing the SEM of the BAL fraction in each bin.

According to Figure \ref{fig:Figure 5}, it appears as if the BAL fraction increases as a function of luminosity. It is also apparent that the BAL fractions of the GNIRS-DQS sources are generally consistent with those of their parent SDSS sample. This trend of increasing BAL fraction as a function of luminosity is also broadly consistent with the results of \citet{Ganguly07b}. A different approach employed by \citet{Allen11} for SDSS quasars reports intrinsic BAL quasar fractions which are somewhat larger than the values we derive from the \citet{Lyke20} sample at higher luminosities. The \citet{Allen11} data can be considered as conservative upper limits to the BAL quasar fractions for luminous quasars. As indicated by \citet{Laor02} and \citet{Ganguly08}, quasars with low luminosity (\hbox{$\lambda L_{\lambda} \lesssim 10^{44}$} erg s$^{-1}$) are not capable of attaining the requisite high maximum velocity of the outflow necessary for the manifestation of the BAL phenomenon according to the W91 definition. This trend is consistent with the data in Figure \ref{fig:Figure 5}.

Instead of the increased ability to launch outflows as luminosity increases, there are alternative explanations to the observed trend of increased BAL fraction as a function of luminosity. For example, \citet{Yuan03} argue that the trend of increasing BAL fraction with increasing luminosity could be a consequence of low-luminosity sources having a shortage of gas supply and hence lower Eddington ratios. Therefore, sources that have the largest supply of cold gas (at higher redshifts), have the largest fueling and outflow rates and exhibit BAL troughs. Another explanation for this trend may be related to low-luminosity quasars typically having low S/N spectra, in which a BAL trough is less likely to be identified as such (e.g., \citealt{Allen11}, R20).

At lower redshifts ($z \lesssim 1.5$), the characterization of BAL quasars remains unclear primarily due to the scarcity of available rest-frame UV data as the majority of BAL quasar studies lack a statistically significant number of sources. For example, \citet{Brandt00} reports five BAL quasars out of a sample of 87 sources. Similarly, the \citet{Ganguly07a} sample includes two BAL quasars out of 14 sources, and the \citet{Matthews17} sample has 58 BAL quasars out of 16,742 sources, though their sample consists exclusively of LoBAL quasars. Moreover, since luminous quasars lie preferentially at higher redshifts and the BAL quasar fraction decreases toward lower luminosities, the census of nearby BAL quasar targets is quite limited.

Another factor that may contribute to the observation of BAL troughs in quasar spectra is orientation. The outflows in BAL quasars may not be directed along our line of sight, which might potentially account for the observed BAL fraction of 10-15\% at viewing angles less than $\sim30$\textdegree\ relative to the plane of the disk within the larger population of quasars (e.g., \citealt{Zhou06}; \citealt{DiPompeo12}). An additional complication that arises is that quasars viewed at angles above 45\textdegree\ (more edge-on) are likely obscured by dust (e.g., \citealt{Barthel89}). 

\citet{Nair22} observe that the BAL quasar sample at high orientation angles is almost double the BAL quasar fraction at low orientation angles ($\sim 40\%$ compared to $\sim 20\%$), pointing to the possible existence of polar BAL quasars. \citet{Kunert-Bajraszewska15} also found that stronger C~{\sc iv}  absorption is associated with lower values of the radio-loudness parameter, and thus with high orientation angles for BAL quasars. This trend is consistent with our GNIRS-DQS sample of 65 BAL quasars, only one of which is also RL (see Section \ref{sec:sample selection}), constituting a considerably lower fraction of RL sources among the general quasar population. This observation supports the idea that BAL quasars tend to occupy large orientation angles. Similar findings by \citet{DiPompeo12} indicate a general trend suggesting that BAL quasar viewing angles extend about 10\textdegree\ farther from the radio jet axis.

Figure \ref{fig:Figure 5} indicates that a certain threshold luminosity, \hbox{$\lambda L_{2500} \gtrsim 10^{45}$ erg s$^{-1}$}, is a necessary condition for the production of outflows. While high luminosity is required, it is clearly insufficient to guarantee the appearance of a BAL trough. Even in the most luminous quasars, the BAL fraction remains below \hbox{$\sim$ 40\%} (see Figure \ref{fig:Figure 5}, but, see also Figure 28 of \citealt{Allen11} for a somewhat larger intrinsic BAL  fraction). This trend can be explained by a limited BAL covering fraction observed at preferential orientation angles. All this suggests that orientation plays a significant role in the appearance of BAL troughs in quasar spectra (see also e.g., \citealt{Filiz13}, \citeyear{2014ApJ...791...88F}).

By design, GNIRS-DQS encompasses only high luminosity, and consequently high $M_{\rm BH}$ values, providing limited insights into the properties of BAL quasars across the broader quasar parameter space (see M21; Figure 5). Extending GNIRS-DQS to lower luminosities would allow inclusion of distant sources with lower $M_{\rm BH}$ values, enabling one to disentangle the roles that luminosity, $M_{\rm BH}$, and orientation play in the appearance of BAL troughs in quasar spectra. This goal could be achieved most efficiently by employing NIR spectrographs on future large observatories such as the European Extremely Large Telescope (e.g., \citealt{Thatte16}), the Giant Magellan Telescope (e.g., \citealt{Jaffe16}), or the Thirty Meter Telescope (e.g., \citealt{Larkin16}).

\raggedbottom

\section{Summary and Conclusions} \label{sec:conclusion}

In this work, we conduct a comparative analysis of the rest-frame optical properties of 65 BAL quasars and 195 non-BAL quasars to gain insights into the appearance of BAL features in quasar spectra. All the quasars were drawn from GNIRS-DQS, a flux-limited uniform sample of luminous sources at $1.55 \lesssim z \lesssim 3.50$ that have spectroscopic data of the H$\beta$ region. Our BAL quasar sample, which constitutes a BAL quasar fraction of 25\%, is the largest uniform sample of such sources having rest-frame optical spectral properties. We perform careful comparisons between BAL quasars and non-BAL quasars based on velocity offsets from systemic redshifts and emission-line properties, such as EW, FWHM, and $R_{\rm Fe~{II}}$. We also explore correlations between BAL trough properties and the H$\beta$-based supermassive black hole masses and normalized accretion rates.

We find that, in spite of the differences between luminous BAL and non-BAL quasars in their rest-frame UV spectra, they are generally indistinguishable in their rest-frame optical spectra. This result is broadly consistent with the idea that the BALs observed in the rest-frame UV spectra of luminous quasars are the result of clumpy, outflowing gas along the line of sight (e.g., W91; \citealt{Yong18}). Specifically, all luminous quasars have the potential to be seen as BAL quasars, but the appearance of BAL features may depend on the orientation of the outflow with respect to our line-of-sight. We do not find any significant correlations between BAL trough properties and either $M_{\rm BH}$ or $L/L_{\rm Edd}$ in our sample and, overall, the velocity offsets from systemic redshifts of our BAL quasars are similar to their non-BAL counterparts. A future extension of the high-redshift sample to considerably lower luminosities may allow one to explore the contributions of luminosity, $M_{\rm BH}$, and orientation to the quasar BAL phenomenon.

\raggedbottom
\begin{acknowledgments}
This work is supported by National Science Foundation (NSF) grants AST-1815281 (H. A., B. M. M., C. D., O. S.) and AST-1815645 (M. S. B., A. D. M., J. N. M.). We thank an anonymous reviewer for valuable comments that helped improve this manuscript. This research has made use of the NASA/IPAC Extra-galactic Database (NED), which is operated by the Jet Propulsion Laboratory, California Institute of Technology, under contract with the National Aeronautics and Space Administration. T. H. acknowledges support from a pre-doctoral program at the Flatiron Institute. Research at the Flatiron Institute is supported by the Simons Foundation.
\end{acknowledgments}

\software{MATLAB and Statistics Toolbox \citep{MATLAB},
        matplotlib \citep{Hunter07},  
          numpy \citep{van11}; \citep{Harris20}, 
          pandas \citep{Mckinney10},
          scipy \citep{Virtanen20},
          scikit-learn \citep{Pedregosa11}
          }

\bibliographystyle{aasjournal}
\bibliography{references}{}

\begin{thebibliography}{}
\expandafter\ifx\csname natexlab\endcsname\relax\def\natexlab#1{#1}\fi
\providecommand{\url}[1]{\href{#1}{#1}}
\providecommand{\dodoi}[1]{doi:~\href{http://doi.org/#1}{\nolinkurl{#1}}}
\providecommand{\doeprint}[1]{\href{http://ascl.net/#1}{\nolinkurl{http://ascl.net/#1}}}
\providecommand{\doarXiv}[1]{\href{https://arxiv.org/abs/#1}{\nolinkurl{https://arxiv.org/abs/#1}}}

\bibitem[{{Abazajian} {et~al.}(2009){Abazajian}, {Adelman-McCarthy},
  {Ag{\"u}eros}, {Allam}, {Allende Prieto}, {An}, {Anderson}, {Anderson},
  {Annis}, {Bahcall}, {Bailer-Jones}, {Barentine}, {Bassett}, {Becker},
  {Beers}, {Bell}, {Belokurov}, {Berlind}, {Berman}, {Bernardi}, {Bickerton},
  {Bizyaev}, {Blakeslee}, {Blanton}, {Bochanski}, {Boroski}, {Brewington},
  {Brinchmann}, {Brinkmann}, {Brunner}, {Budav{\'a}ri}, {Carey}, {Carliles},
  {Carr}, {Castander}, {Cinabro}, {Connolly}, {Csabai}, {Cunha}, {Czarapata},
  {Davenport}, {de Haas}, {Dilday}, {Doi}, {Eisenstein}, {Evans}, {Evans},
  {Fan}, {Friedman}, {Frieman}, {Fukugita}, {G{\"a}nsicke}, {Gates},
  {Gillespie}, {Gilmore}, {Gonzalez}, {Gonzalez}, {Grebel}, {Gunn},
  {Gy{\"o}ry}, {Hall}, {Harding}, {Harris}, {Harvanek}, {Hawley}, {Hayes},
  {Heckman}, {Hendry}, {Hennessy}, {Hindsley}, {Hoblitt}, {Hogan}, {Hogg},
  {Holtzman}, {Hyde}, {Ichikawa}, {Ichikawa}, {Im}, {Ivezi{\'c}}, {Jester},
  {Jiang}, {Johnson}, {Jorgensen}, {Juri{\'c}}, {Kent}, {Kessler}, {Kleinman},
  {Knapp}, {Konishi}, {Kron}, {Krzesinski}, {Kuropatkin}, {Lampeitl},
  {Lebedeva}, {Lee}, {Lee}, {French Leger}, {L{\'e}pine}, {Li}, {Lima}, {Lin},
  {Long}, {Loomis}, {Loveday}, {Lupton}, {Magnier}, {Malanushenko},
  {Malanushenko}, {Mandelbaum}, {Margon}, {Marriner}, {Mart{\'\i}nez-Delgado},
  {Matsubara}, {McGehee}, {McKay}, {Meiksin}, {Morrison}, {Mullally}, {Munn},
  {Murphy}, {Nash}, {Nebot}, {Neilsen}, {Newberg}, {Newman}, {Nichol},
  {Nicinski}, {Nieto-Santisteban}, {Nitta}, {Okamura}, {Oravetz}, {Ostriker},
  {Owen}, {Padmanabhan}, {Pan}, {Park}, {Pauls}, {Peoples}, {Percival}, {Pier},
  {Pope}, {Pourbaix}, {Price}, {Purger}, {Quinn}, {Raddick}, {Re Fiorentin},
  {Richards}, {Richmond}, {Riess}, {Rix}, {Rockosi}, {Sako}, {Schlegel},
  {Schneider}, {Scholz}, {Schreiber}, {Schwope}, {Seljak}, {Sesar}, {Sheldon},
  {Shimasaku}, {Sibley}, {Simmons}, {Sivarani}, {Allyn Smith}, {Smith},
  {Smol{\v{c}}i{\'c}}, {Snedden}, {Stebbins}, {Steinmetz}, {Stoughton},
  {Strauss}, {SubbaRao}, {Suto}, {Szalay}, {Szapudi}, {Szkody}, {Tanaka},
  {Tegmark}, {Teodoro}, {Thakar}, {Tremonti}, {Tucker}, {Uomoto}, {Vanden
  Berk}, {Vandenberg}, {Vidrih}, {Vogeley}, {Voges}, {Vogt}, {Wadadekar},
  {Watters}, {Weinberg}, {West}, {White}, {Wilhite}, {Wonders}, {Yanny},
  {Yocum}, {York}, {Zehavi}, {Zibetti}, \& {Zucker}}]{Abazajian09}
{Abazajian}, K.~N., {Adelman-McCarthy}, J.~K., {Ag{\"u}eros}, M.~A., {et~al.}
  2009, \apjs, 182, 543

\bibitem[{{Allen} {et~al.}(2011){Allen}, {Hewett}, {Maddox}, {Richards}, \&
  {Belokurov}}]{Allen11}
{Allen}, J.~T., {Hewett}, P.~C., {Maddox}, N., {Richards}, G.~T., \&
  {Belokurov}, V. 2011, \mnras, 410, 860

\bibitem[{{Baldwin}(1977)}]{Baldwin77}
{Baldwin}, J.~A. 1977, \apj, 214, 679

\bibitem[{{Barthel}(1989)}]{Barthel89}
{Barthel}, P.~D. 1989, \apj, 336, 606

\bibitem[{{Baskin} {et~al.}(2013){Baskin}, {Laor}, \& {Hamann}}]{Baskin13}
{Baskin}, A., {Laor}, A., \& {Hamann}, F. 2013, \mnras, 432, 1525

\bibitem[{{Becker} {et~al.}(1997){Becker}, {Gregg}, {Hook}, {McMahon}, {White},
  \& {Helfand}}]{Becker97}
{Becker}, R.~H., {Gregg}, M.~D., {Hook}, I.~M., {et~al.} 1997, \apjl, 479, L93

\bibitem[{{Becker} {et~al.}(2001){Becker}, {White}, {Gregg},
  {Laurent-Muehleisen}, {Brotherton}, {Impey}, {Chaffee}, {Richards},
  {Helfand}, {Lacy}, {Courbin}, \& {Proctor}}]{Becker01}
{Becker}, R.~H., {White}, R.~L., {Gregg}, M.~D., {et~al.} 2001, VizieR Online
  Data Catalog, J/ApJS/135/227

\bibitem[{{Begelman} {et~al.}(2006){Begelman}, {Volonteri}, \&
  {Rees}}]{Begelman06}
{Begelman}, M.~C., {Volonteri}, M., \& {Rees}, M.~J. 2006, \mnras, 370, 289

\bibitem[{{Bieri} {et~al.}(2017){Bieri}, {Dubois}, {Rosdahl}, {Wagner}, {Silk},
  \& {Mamon}}]{Bieri17}
{Bieri}, R., {Dubois}, Y., {Rosdahl}, J., {et~al.} 2017, \mnras, 464, 1854

\bibitem[{{Boroson}(2005)}]{Boroson05}
{Boroson}, T. 2005, \aj, 130, 381

\bibitem[{{Boroson} \& {Green}(1992)}]{Boroson92}
{Boroson}, T.~A., \& {Green}, R.~F. 1992, \apjs, 80, 109

\bibitem[{{Brandt} {et~al.}(2000){Brandt}, {Laor}, \& {Wills}}]{Brandt00}
{Brandt}, W.~N., {Laor}, A., \& {Wills}, B.~J. 2000, \apj, 528, 637

\bibitem[{{Brotherton} {et~al.}(2006){Brotherton}, {De Breuck}, \&
  {Schaefer}}]{Brotherton06}
{Brotherton}, M.~S., {De Breuck}, C., \& {Schaefer}, J.~J. 2006, \mnras, 372,
  L58

\bibitem[{{Brotherton} {et~al.}(1998){Brotherton}, {van Breugel}, {Smith},
  {Boyle}, {Shanks}, {Croom}, {Miller}, \& {Becker}}]{Brotherton98}
{Brotherton}, M.~S., {van Breugel}, W., {Smith}, R.~J., {et~al.} 1998, \apjl,
  505, L7

\bibitem[{{Calistro Rivera} {et~al.}(2023){Calistro Rivera}, {Alexander},
  {Harrison}, {Fawcett}, {Best}, {Williams}, {Hardcastle}, {Rosario}, {Smith},
  {Arnaudova}, {Escott}, {G{\"u}rkan}, {Kondapally}, {Miley}, {Morabito},
  {Petley}, {Prandoni}, {R{\"o}ttgering}, \& {Yue}}]{Calisto23}
{Calistro Rivera}, G., {Alexander}, D.~M., {Harrison}, C.~M., {et~al.} 2023,
  arXiv e-prints

\bibitem[{{Cattaneo} {et~al.}(2005){Cattaneo}, {Blaizot}, {Devriendt}, \&
  {Guiderdoni}}]{Cattaneo05}
{Cattaneo}, A., {Blaizot}, J., {Devriendt}, J., \& {Guiderdoni}, B. 2005,
  \mnras, 364, 407

\bibitem[{{Di Matteo} {et~al.}(2005){Di Matteo}, {Springel}, \&
  {Hernquist}}]{DiMatteo05}
{Di Matteo}, T., {Springel}, V., \& {Hernquist}, L. 2005, \nat, 433, 604

\bibitem[{{Dietrich} {et~al.}(2002){Dietrich}, {Hamann}, {Shields},
  {Constantin}, {Vestergaard}, {Chaffee}, {Foltz}, \&
  {Junkkarinen}}]{Dietrich02}
{Dietrich}, M., {Hamann}, F., {Shields}, J.~C., {et~al.} 2002, \apj, 581, 912

\bibitem[{{DiPompeo} {et~al.}(2012){DiPompeo}, {Brotherton}, \& {De
  Breuck}}]{DiPompeo12}
{DiPompeo}, M.~A., {Brotherton}, M.~S., \& {De Breuck}, C. 2012, \apj, 752, 6

\bibitem[{{DiPompeo} {et~al.}(2014){DiPompeo}, {Myers}, {Brotherton}, {Runnoe},
  \& {Green}}]{DiPompeo14}
{DiPompeo}, M.~A., {Myers}, A.~D., {Brotherton}, M.~S., {Runnoe}, J.~C., \&
  {Green}, R.~F. 2014, \apj, 787, 73

\bibitem[{{Dix} {et~al.}(2023){Dix}, {Matthews}, {Shemmer}, {Brotherton},
  {Myers}, {Andruchow}, {Brandt}, {Ferrero}, {Green}, {Lira}, {Plotkin},
  {Richards}, \& {Schneider}}]{Dix23}
{Dix}, C., {Matthews}, B., {Shemmer}, O., {et~al.} 2023, \apj, 950, 96

\bibitem[{{Du} \& {Wang}(2019)}]{Du19}
{Du}, P., \& {Wang}, J.-M. 2019, \apj, 886, 42

\bibitem[{{Elvis}(2000)}]{Elvis2000}
{Elvis}, M. 2000, \apj, 545, 63

\bibitem[{{Filiz Ak} {et~al.}(2013){Filiz Ak}, {Brandt}, {Hall}, {Schneider},
  {Anderson}, {Hamann}, {Lundgren}, {Myers}, {P{\^a}ris}, {Petitjean}, {Ross},
  {Shen}, \& {York}}]{Filiz13}
{Filiz Ak}, N., {Brandt}, W.~N., {Hall}, P.~B., {et~al.} 2013, \apj, 777, 168

\bibitem[{{{Filiz Ak}, N. and {Brandt}, W.~N. and {Hall}, P.~B. and
  {Schneider}, D.~P. and {Trump}, J.~R. and {Anderson}, S.~F. and {Hamann}, F.
  and {Myers}, Adam D. and {P{\^a}ris}, I. and {Petitjean}, P. and {Ross},
  Nicholas P. and {Shen}, Yue and {York}, Don}(2014)}]{2014ApJ...791...88F}
{{Filiz Ak}, N. and {Brandt}, W.~N. and {Hall}, P.~B. and {Schneider}, D.~P.
  and {Trump}, J.~R. and {Anderson}, S.~F. and {Hamann}, F. and {Myers}, Adam
  D. and {P{\^a}ris}, I. and {Petitjean}, P. and {Ross}, Nicholas P. and
  {Shen}, Yue and {York}, Don}. 2014, \apj, 791, 88

\bibitem[{{Gallagher} {et~al.}(2002){Gallagher}, {Brandt}, {Chartas}, \&
  {Garmire}}]{Gall02}
{Gallagher}, S.~C., {Brandt}, W.~N., {Chartas}, G., \& {Garmire}, G.~P. 2002,
  \apj, 567, 37

\bibitem[{{Gallagher} {et~al.}(2006){Gallagher}, {Brandt}, {Chartas},
  {Priddey}, {Garmire}, \& {Sambruna}}]{Gall06}
{Gallagher}, S.~C., {Brandt}, W.~N., {Chartas}, G., {et~al.} 2006, \apj, 644,
  709

\bibitem[{{Gallagher} {et~al.}(2007){Gallagher}, {Hines}, {Blaylock},
  {Priddey}, {Brandt}, \& {Egami}}]{Gall07}
{Gallagher}, S.~C., {Hines}, D.~C., {Blaylock}, M., {et~al.} 2007, \apj, 665,
  157

\bibitem[{{Ganguly} \& {Brotherton}(2008)}]{Ganguly08}
{Ganguly}, R., \& {Brotherton}, M.~S. 2008, \apj, 672, 102

\bibitem[{{Ganguly} {et~al.}(2007{\natexlab{a}}){Ganguly}, {Brotherton},
  {Cales}, {Scoggins}, {Shang}, \& {Vestergaard}}]{Ganguly07a}
{Ganguly}, R., {Brotherton}, M.~S., {Cales}, S., {et~al.} 2007{\natexlab{a}},
  \apj, 665, 990

\bibitem[{{Ganguly} {et~al.}(2007{\natexlab{b}}){Ganguly}, {Brotherton},
  {Arav}, {Heap}, {Wisotzki}, {Aldcroft}, {Alloin}, {Behar}, {Canalizo},
  {Crenshaw}, {de Kool}, {Chambers}, {Cecil}, {Chatzichristou}, {Everett},
  {Gabel}, {Gaskell}, {Galliano}, {Green}, {Hall}, {Hines}, {Junkkarinen},
  {Kaastra}, {Kaiser}, {Kazanas}, {Konigl}, {Korista}, {Kriss}, {Laor},
  {Leighly}, {Mathur}, {Ogle}, {Proga}, {Sabra}, {Sivron}, {Snedden}, {Telfer},
  \& {Vestergaard}}]{Ganguly07b}
{Ganguly}, R., {Brotherton}, M.~S., {Arav}, N., {et~al.} 2007{\natexlab{b}},
  \aj, 133, 479

\bibitem[{{Garc{\'\i}a} {et~al.}(2023){Garc{\'\i}a}, {Martini},
  {Gonzalez-Morales}, {Font-Ribera}, {Herrera-Alcantar}, {Aguilar}, {Ahlen},
  {Brooks}, {de la Macorra}, {Doel}, {Forero-Romero}, {Guy}, {Kisner},
  {Landriau}, {Miquel}, {Moustakas}, {Nie}, {Poppett}, {Tarl{\'e}}, \&
  {Zhou}}]{Garcia23}
{Garc{\'\i}a}, L.~{\'A}., {Martini}, P., {Gonzalez-Morales}, A.~X., {et~al.}
  2023, \mnras, 526, 4848

\bibitem[{{Gibson} {et~al.}(2009){Gibson}, {Jiang}, {Brandt}, {Hall}, {Shen},
  {Wu}, {Anderson}, {Schneider}, {Vanden Berk}, {Gallagher}, {Fan}, \&
  {York}}]{Gibson09}
{Gibson}, R.~R., {Jiang}, L., {Brandt}, W.~N., {et~al.} 2009, \apj, 692, 758

\bibitem[{{Giustini} \& {Proga}(2019)}]{Giustini19}
{Giustini}, M., \& {Proga}, D. 2019, \aap, 630, A94

\bibitem[{{Gregg} {et~al.}(2006){Gregg}, {Becker}, \& {de Vries}}]{Gregg06}
{Gregg}, M.~D., {Becker}, R.~H., \& {de Vries}, W. 2006, \apj, 641, 210

\bibitem[{{Ha} {et~al.}(2023){Ha}, {Dix}, {Matthews}, {Shemmer}, {Brotherton},
  {Myers}, {Richards}, {Maithil}, {Anderson}, {Brandt}, {Diamond-Stanic},
  {Fan}, {Gallagher}, {Green}, {Lira}, {Luo}, {Netzer}, {Plotkin}, {Runnoe},
  {Schneider}, {Strauss}, {Trakhtenbrot}, \& {Wu}}]{Ha23}
{Ha}, T., {Dix}, C., {Matthews}, B.~M., {et~al.} 2023, \apj, 950, 97

\bibitem[{{Hall} {et~al.}(2002){Hall}, {Anderson}, {Strauss}, {York},
  {Richards}, {Fan}, {Knapp}, {Schneider}, {Vanden Berk}, {Geballe}, {Bauer},
  {Becker}, {Davis}, {Rix}, {Nichol}, {Bahcall}, {Brinkmann}, {Brunner},
  {Connolly}, {Csabai}, {Doi}, {Fukugita}, {Gunn}, {Haiman}, {Harvanek},
  {Heckman}, {Hennessy}, {Inada}, {Ivezi{\'c}}, {Johnston}, {Kleinman},
  {Krolik}, {Krzesinski}, {Kunszt}, {Lamb}, {Long}, {Lupton}, {Miknaitis},
  {Munn}, {Narayanan}, {Neilsen}, {Newman}, {Nitta}, {Okamura}, {Pentericci},
  {Pier}, {Schlegel}, {Snedden}, {Szalay}, {Thakar}, {Tsvetanov}, {White}, \&
  {Zheng}}]{Hall02}
{Hall}, P.~B., {Anderson}, S.~F., {Strauss}, M.~A., {et~al.} 2002, \apjs, 141,
  267

\bibitem[{{Harris} {et~al.}(2020){Harris}, {Millman}, {van der Walt},
  {Gommers}, {Virtanen}, {Cournapeau}, {Wieser}, {Taylor}, {Berg}, {Smith},
  {Kern}, {Picus}, {Hoyer}, {van Kerkwijk}, {Brett}, {Haldane}, {del R{\'\i}o},
  {Wiebe}, {Peterson}, {G{\'e}rard-Marchant}, {Sheppard}, {Reddy}, {Weckesser},
  {Abbasi}, {Gohlke}, \& {Oliphant}}]{Harris20}
{Harris}, C.~R., {Millman}, K.~J., {van der Walt}, S.~J., {et~al.} 2020, \nat,
  585, 357

\bibitem[{{Hopkins} {et~al.}(2005){Hopkins}, {Hernquist}, {Cox}, {Di Matteo},
  {Martini}, {Robertson}, \& {Springel}}]{Hopkins05}
{Hopkins}, P.~F., {Hernquist}, L., {Cox}, T.~J., {et~al.} 2005, \apj, 630, 705

\bibitem[{Hu {et~al.}(2006)Hu, Shen, Lou, \& Zhang}]{Hu06}
Hu, J., Shen, Y., Lou, Y.-Q., \& Zhang, S. 2006, \mnras, 365, 345

\bibitem[{{Hunter}(2007)}]{Hunter07}
{Hunter}, J.~D. 2007, Computing in Science and Engineering, 9, 90

\bibitem[{{Jaffe} {et~al.}(2016){Jaffe}, {Barnes}, {Brooks}, {Lee}, {Mace},
  {Pak}, {Park}, \& {Park}}]{Jaffe16}
{Jaffe}, D.~T., {Barnes}, S., {Brooks}, C., {et~al.} 2016, in SPIE, Vol. 9908,
  Ground-based and Airborne Instrumentation for Astronomy VI, ed. C.~J.
  {Evans}, L.~{Simard}, \& H.~{Takami}, 990821

\bibitem[{{Kellermann} {et~al.}(1989){Kellermann}, {Sramek}, {Schmidt},
  {Shaffer}, \& {Green}}]{Kellermann89}
{Kellermann}, K.~I., {Sramek}, R., {Schmidt}, M., {Shaffer}, D.~B., \& {Green},
  R. 1989, \aj, 98, 1195

\bibitem[{{Knigge} {et~al.}(2008){Knigge}, {Scaringi}, {Goad}, \&
  {Cottis}}]{Knigge08}
{Knigge}, C., {Scaringi}, S., {Goad}, M.~R., \& {Cottis}, C.~E. 2008, \mnras,
  386, 1426

\bibitem[{{Kunert-Bajraszewska} {et~al.}(2015){Kunert-Bajraszewska},
  {Ceg{\l}owski}, {Katarzy{\'n}ski}, \&
  {Roskowi{\'n}ski}}]{Kunert-Bajraszewska15}
{Kunert-Bajraszewska}, M., {Ceg{\l}owski}, M., {Katarzy{\'n}ski}, K., \&
  {Roskowi{\'n}ski}, C. 2015, \aap, 579, A109

\bibitem[{{Laor} \& {Brandt}(2002)}]{Laor02}
{Laor}, A., \& {Brandt}, W.~N. 2002, \apj, 569, 641

\bibitem[{{Larkin} {et~al.}(2016){Larkin}, {Moore}, {Wright}, {Wincentsen},
  {Anderson}, {Chisholm}, {Dekany}, {Dunn}, {Ellerbroek}, {Hayano}, {Phillips},
  {Simard}, {Smith}, {Suzuki}, {Weber}, {Weiss}, \& {Zhang}}]{Larkin16}
{Larkin}, J.~E., {Moore}, A.~M., {Wright}, S.~A., {et~al.} 2016, in SPIE, Vol.
  9908, Ground-based and Airborne Instrumentation for Astronomy VI, ed. C.~J.
  {Evans}, L.~{Simard}, \& H.~{Takami}, 99081W

\bibitem[{{Leighly} {et~al.}(2022){Leighly}, {Choi}, {DeFrancesco}, {Voelker},
  {Terndrup}, {Gallagher}, \& {Richards}}]{Leighly22}
{Leighly}, K.~M., {Choi}, H., {DeFrancesco}, C., {et~al.} 2022, \apj, 935, 92

\bibitem[{{L{\'\i}pari} \& {Terlevich}(2006)}]{Lipari06}
{L{\'\i}pari}, S.~L., \& {Terlevich}, R.~J. 2006, \mnras, 368, 1001

\bibitem[{{Liu} {et~al.}(2018){Liu}, {Luo}, {Brandt}, {Gallagher}, \&
  {Garmire}}]{Liu18}
{Liu}, H., {Luo}, B., {Brandt}, W.~N., {Gallagher}, S.~C., \& {Garmire}, G.~P.
  2018, \apj, 859, 113

\bibitem[{{Luo} {et~al.}(2014){Luo}, {Brandt}, {Alexander}, {Stern}, {Teng},
  {Ar{\'e}valo}, {Bauer}, {Boggs}, {Christensen}, {Comastri}, {Craig},
  {Farrah}, {Gandhi}, {Hailey}, {Harrison}, {Koss}, {Ogle}, {Puccetti}, {Saez},
  {Scott}, {Walton}, \& {Zhang}}]{Luo14}
{Luo}, B., {Brandt}, W.~N., {Alexander}, D.~M., {et~al.} 2014, \apj, 794, 70

\bibitem[{{Lyke} {et~al.}(2020){Lyke}, {Higley}, {McLane}, {Schurhammer},
  {Myers}, {Ross}, {Dawson}, {Chabanier}, {Martini}, {Busca}, {Du Mas Des
  Bourboux}, {Salvato}, {Streblyanska}, {Zarrouk}, {Burtin}, {Anderson},
  {Bautista}, {Bizyaev}, {Brandt}, {Brinkmann}, {Brownstein}, {Comparat},
  {Green}, {de La Macorra}, {Munoz Gutierrez}, {Hou}, {Newman}, \&
  {Palanque}}]{Lyke20}
{Lyke}, B.~W., {Higley}, A.~N., {McLane}, J.~N., {et~al.} 2020, VizieR Online
  Data Catalog, VII/289

\bibitem[{{Magorrian} {et~al.}(1998){Magorrian}, {Tremaine}, {Richstone},
  {Bender}, {Bower}, {Dressler}, {Faber}, {Gebhardt}, {Green}, {Grillmair},
  {Kormendy}, \& {Lauer}}]{Magorrian98}
{Magorrian}, J., {Tremaine}, S., {Richstone}, D., {et~al.} 1998, \aj, 115, 2285

\bibitem[{{Maithil} {et~al.}(2022){Maithil}, {Brotherton}, {Shemmer}, {Du},
  {Wang}, {Myers}, {McLane}, {Dix}, \& {Matthews}}]{Maithil22}
{Maithil}, J., {Brotherton}, M.~S., {Shemmer}, O., {et~al.} 2022, \mnras, 515,
  491

\bibitem[{{Marconi} \& {Hunt}(2003)}]{Marconi03}
{Marconi}, A., \& {Hunt}, L.~K. 2003, \apjl, 589, L21

\bibitem[{MATLAB(2022)}]{MATLAB}
MATLAB. 2022, version 9.12.0 (R2022a),  Natick, Massachusetts: The MathWorks
  Inc.

\bibitem[{{Matthews} {et~al.}(2021){Matthews}, {Shemmer}, {Dix}, {Brotherton},
  {Myers}, {Andruchow}, {Brandt}, {Ferrero}, {Gallagher}, {Green}, {Lira},
  {Plotkin}, {Richards}, {Runnoe}, {Schneider}, {Shen}, {Strauss}, \&
  {Wills}}]{M21}
{Matthews}, B.~M., {Shemmer}, O., {Dix}, C., {et~al.} 2021, \apjs, 252, 15

\bibitem[{{Matthews} {et~al.}(2023){Matthews}, {Dix}, {Shemmer}, {Brotherton},
  {Myers}, {Andruchow}, {Brandt}, {Gallagher}, {Green}, {Lira}, {McLane},
  {Plotkin}, {Richards}, {Runnoe}, {Schneider}, \& {Strauss}}]{M23}
{Matthews}, B.~M., {Dix}, C., {Shemmer}, O., {et~al.} 2023, \apj, 950, 95

\bibitem[{{Matthews} {et~al.}(2017){Matthews}, {Knigge}, \&
  {Long}}]{Matthews17}
{Matthews}, J.~H., {Knigge}, C., \& {Long}, K.~S. 2017, \mnras, 467, 2571

\bibitem[{{Morabito} {et~al.}(2014){Morabito}, {Dai}, {Leighly}, {Sivakoff}, \&
  {Shankar}}]{Morabito14}
{Morabito}, L.~K., {Dai}, X., {Leighly}, K.~M., {Sivakoff}, G.~R., \&
  {Shankar}, F. 2014, \apj, 786, 58

\bibitem[{{Murray} {et~al.}(1995){Murray}, {Chiang}, {Grossman}, \&
  {Voit}}]{Murray95}
{Murray}, N., {Chiang}, J., {Grossman}, S.~A., \& {Voit}, G.~M. 1995, \apj,
  451, 498

\bibitem[{{Nair} \& {Vivek}(2022)}]{Nair22}
{Nair}, A., \& {Vivek}, M. 2022, \mnras, 511, 4946

\bibitem[{{Netzer} {et~al.}(2004){Netzer}, {Shemmer}, {Maiolino}, {Oliva},
  {Croom}, {Corbett}, \& {di Fabrizio}}]{Netzer04}
{Netzer}, H., {Shemmer}, O., {Maiolino}, R., {et~al.} 2004, \apj, 614, 558

\bibitem[{{Netzer} \& {Trakhtenbrot}(2007)}]{Netzer07}
{Netzer}, H., \& {Trakhtenbrot}, B. 2007, \apj, 654, 754

\bibitem[{{Nguyen} {et~al.}(2020){Nguyen}, {Lira}, {Trakhtenbrot}, {Netzer},
  {Cicone}, {Maiolino}, \& {Shemmer}}]{Nguyen20}
{Nguyen}, N.~H., {Lira}, P., {Trakhtenbrot}, B., {et~al.} 2020, \apj, 895, 74

\bibitem[{{Ogle} {et~al.}(1999){Ogle}, {Cohen}, {Miller}, {Tran}, {Goodrich},
  \& {Martel}}]{Ogle91}
{Ogle}, P.~M., {Cohen}, M.~H., {Miller}, J.~S., {et~al.} 1999, \apjs, 125, 1

\bibitem[{{Paris} {et~al.}(2017){Paris}, {Petitjean}, {Ross}, {Myers},
  {Aubourg}, {Streblyanska}, {Bailey}, {Armengaud}, {Palanque-Delabrouille},
  {Yeche}, {Hamann}, {Strauss}, {Albareti}, {Bovy}, {Bizyaev}, {Brandt},
  {Brusa}, {Buchner}, {Comparat}, {Croft}, {Dwelly}, {Fan}, {Font-Ribera},
  {Ge}, {Georgakakis}, {Hall}, {Jiang}, {Kinemuchi}, {Malanushenko},
  {Malanushenko}, {McMahon}, {Menzel}, {Merloni}, {Nandra}, {Noterdaeme},
  {Oravetz}, {Pan}, {Pieri}, {Prada}, {Salvato}, {Schlegel}, {Schneider},
  {Simmons}, {Viel}, {Weinberg}, \& {Zhu}}]{Paris17}
{Paris}, I., {Petitjean}, P., {Ross}, N.~P., {et~al.} 2017, VizieR Online Data
  Catalog, VII/279

\bibitem[{{P{\^a}ris} {et~al.}(2018){P{\^a}ris}, {Petitjean}, {Aubourg},
  {Myers}, {Streblyanska}, {Lyke}, {Anderson}, {Armengaud}, {Bautista},
  {Blanton}, {Blomqvist}, {Brinkmann}, {Brownstein}, {Brandt}, {Burtin},
  {Dawson}, {de la Torre}, {Georgakakis}, {Gil-Mar{\'\i}n}, {Green}, {Hall},
  {Kneib}, {LaMassa}, {Le Goff}, {MacLeod}, {Mariappan}, {McGreer}, {Merloni},
  {Noterdaeme}, {Palanque-Delabrouille}, {Percival}, {Ross}, {Rossi},
  {Schneider}, {Seo}, {Tojeiro}, {Weaver}, {Weijmans}, {Y{\`e}che}, {Zarrouk},
  \& {Zhao}}]{Paris18}
{P{\^a}ris}, I., {Petitjean}, P., {Aubourg}, {\'E}., {et~al.} 2018, \aap, 613,
  A51

\bibitem[{{Pedregosa} {et~al.}(2011){Pedregosa}, {Varoquaux}, {Gramfort},
  {Michel}, {Thirion}, {Grisel}, {Blondel}, {M{\"u}ller}, {Nothman}, {Louppe},
  {Prettenhofer}, {Weiss}, {Dubourg}, {Vanderplas}, {Passos}, {Cournapeau},
  {Brucher}, {Perrot}, \& {Duchesnay}}]{Pedregosa11}
{Pedregosa}, F., {Varoquaux}, G., {Gramfort}, A., {et~al.} 2011, Journal of
  Machine Learning Research, 12, 2825

\bibitem[{{Petley} {et~al.}(2024){Petley}, {Morabito}, {Rankine}, {Richards},
  {Thomas}, {Alexander}, {Fawcett}, {Rivera}, {Prandoni}, {Best}, \&
  {Kolwa}}]{Petley24}
{Petley}, J.~W., {Morabito}, L.~K., {Rankine}, A.~L., {et~al.} 2024, \mnras

\bibitem[{{Quera-Bofarull} {et~al.}(2023){Quera-Bofarull}, {Done}, {Lacey},
  {Nomura}, \& {Ohsuga}}]{Quera23}
{Quera-Bofarull}, A., {Done}, C., {Lacey}, C.~G., {Nomura}, M., \& {Ohsuga}, K.
  2023, \mnras, 518, 2693

\bibitem[{{Rankine} {et~al.}(2020){Rankine}, {Hewett}, {Banerji}, \&
  {Richards}}]{Rankine20}
{Rankine}, A.~L., {Hewett}, P.~C., {Banerji}, M., \& {Richards}, G.~T. 2020,
  \mnras, 492, 4553

\bibitem[{{Reichard} {et~al.}(2003){Reichard}, {Richards}, {Hall}, {Schneider},
  {Vanden Berk}, {Fan}, {York}, {Knapp}, \& {Brinkmann}}]{Reichard03}
{Reichard}, T.~A., {Richards}, G.~T., {Hall}, P.~B., {et~al.} 2003, \aj, 126,
  2594

\bibitem[{{Richards} {et~al.}(2021){Richards}, {McCaffrey}, {Kimball},
  {Rankine}, {Matthews}, {Hewett}, \& {Rivera}}]{Mcaffrey21}
{Richards}, G.~T., {McCaffrey}, T.~V., {Kimball}, A., {et~al.} 2021, \aj, 162,
  270

\bibitem[{{Richards} {et~al.}(2002){Richards}, {Vanden Berk}, {Reichard},
  {Hall}, {Schneider}, {SubbaRao}, {Thakar}, \& {York}}]{Richards02}
{Richards}, G.~T., {Vanden Berk}, D.~E., {Reichard}, T.~A., {et~al.} 2002, \aj,
  124, 1

\bibitem[{{Richards} {et~al.}(2011){Richards}, {Kruczek}, {Gallagher}, {Hall},
  {Hewett}, {Leighly}, {Deo}, {Kratzer}, \& {Shen}}]{Richards11}
{Richards}, G.~T., {Kruczek}, N.~E., {Gallagher}, S.~C., {et~al.} 2011, \aj,
  141, 167

\bibitem[{{Rivera} {et~al.}(2022){Rivera}, {Richards}, {Gallagher},
  {McCaffrey}, {Rankine}, {Hewett}, \& {Shemmer}}]{Rivera22}
{Rivera}, A.~B., {Richards}, G.~T., {Gallagher}, S.~C., {et~al.} 2022, \apj,
  931, 154

\bibitem[{{Runnoe} {et~al.}(2013){Runnoe}, {Shang}, \& {Brotherton}}]{Runnoe13}
{Runnoe}, J.~C., {Shang}, Z., \& {Brotherton}, M.~S. 2013, \mnras, 435, 3251

\bibitem[{{Schmidt} \& {Hines}(1999)}]{Schmidt99}
{Schmidt}, G.~D., \& {Hines}, D.~C. 1999, \apj, 512, 125

\bibitem[{{Schulze} {et~al.}(2017){Schulze}, {Schramm}, {Zuo}, {Wu}, {Urrutia},
  {Kotilainen}, {Reynolds}, {Terao}, {Nagao}, \& {Izumiura}}]{Schulze17}
{Schulze}, A., {Schramm}, M., {Zuo}, W., {et~al.} 2017, \apj, 848, 104

\bibitem[{{Shemmer} \& {Lieber}(2015)}]{Shemmer15}
{Shemmer}, O., \& {Lieber}, S. 2015, \apj, 805, 124

\bibitem[{{Shen} \& {Ho}(2014)}]{Shen&Ho14}
{Shen}, Y., \& {Ho}, L.~C. 2014, \nat, 513, 210

\bibitem[{{Shen} {et~al.}(2016){Shen}, {Brandt}, {Richards}, {Denney},
  {Greene}, {Grier}, {Ho}, {Peterson}, {Petitjean}, {Schneider}, {Tao}, \&
  {Trump}}]{Shen16}
{Shen}, Y., {Brandt}, W.~N., {Richards}, G.~T., {et~al.} 2016, \apj, 831, 7

\bibitem[{{Silk} \& {Rees}(1998)}]{Silk98}
{Silk}, J., \& {Rees}, M.~J. 1998, \aap, 331, L1

\bibitem[{{Spergel} {et~al.}(2007){Spergel}, {Bean}, {Dor{\'e}}, {Nolta},
  {Bennett}, {Dunkley}, {Hinshaw}, {Jarosik}, {Komatsu}, {Page}, {Peiris},
  {Verde}, {Halpern}, {Hill}, {Kogut}, {Limon}, {Meyer}, {Odegard}, {Tucker},
  {Weiland}, {Wollack}, \& {Wright}}]{Spergel07}
{Spergel}, D.~N., {Bean}, R., {Dor{\'e}}, O., {et~al.} 2007, \apjs, 170, 377

\bibitem[{{Sprayberry} \& {Foltz}(1992)}]{Sprayberry92}
{Sprayberry}, D., \& {Foltz}, C.~B. 1992, \apj, 390, 39

\bibitem[{{Stocke} {et~al.}(1992){Stocke}, {Morris}, {Weymann}, \&
  {Foltz}}]{Stocke92}
{Stocke}, J.~T., {Morris}, S.~L., {Weymann}, R.~J., \& {Foltz}, C.~B. 1992,
  \apj, 396, 487

\bibitem[{{Temple} {et~al.}(2023){Temple}, {Matthews}, {Hewett}, {Rankine},
  {Richards}, {Banerji}, {Ferland}, {Knigge}, \& {Stepney}}]{Temple23}
{Temple}, M.~J., {Matthews}, J.~H., {Hewett}, P.~C., {et~al.} 2023, \mnras,
  523, 646

\bibitem[{{Teng} {et~al.}(2014){Teng}, {Brandt}, {Harrison}, {Luo},
  {Alexander}, {Bauer}, {Boggs}, {Christensen}, {Comastri}, {Craig}, {Fabian},
  {Farrah}, {Fiore}, {Gandhi}, {Grefenstette}, {Hailey}, {Hickox}, {Madsen},
  {Ptak}, {Rigby}, {Risaliti}, {Saez}, {Stern}, {Veilleux}, {Walton}, {Wik}, \&
  {Zhang}}]{Teng14}
{Teng}, S.~H., {Brandt}, W.~N., {Harrison}, F.~A., {et~al.} 2014, \apj, 785, 19

\bibitem[{{Thatte} {et~al.}(2016){Thatte}, {Clarke}, {Bryson}, {Shnetler},
  {Tecza}, {Fusco}, {Bacon}, {Richard}, {Mediavilla}, {Neichel}, {Arribas},
  {Garcia-Lorenzo}, {Evans}, {Remillieux}, {El Madi}, {Herreros}, {Melotte},
  {O'Brien}, {Tosh}, {Vernet}, {Hammersley}, {Ives}, {Finger}, {Houghton},
  {Rigopoulou}, {Lynn}, {Allen}, {Zieleniewski}, {Kendrew}, {Ferraro-Wood},
  {P{\'e}contal-Rousset}, {Kosmalski}, {Laurent}, {Loupias}, {Piqueras},
  {Renault}, {Blaizot}, {Daguis{\'e}}, {Migniau}, {Jarno}, {Born}, {Gallie},
  {Montgomery}, {Henry}, {Schwartz}, {Taylor}, {Zins}, {Rodr{\'\i}guez-Ramos},
  {Cagigas}, {Battaglia}, {Rebolo L{\'o}pez}, {Hern{\'a}ndez Su{\'a}rez},
  {Gigante-Ripoll}, {Piqueras L{\'o}pez}, {Villar Martin}, {Correia}, {Pascal},
  {Blanco}, {Vola}, {Epinat}, {Peroux}, {Vigan}, {Dohlen}, {Sauvage}, {Lee},
  {Carlotti}, {Verinaud}, {Morris}, {Myers}, {Reeves}, {Swinbank}, {Calcines},
  \& {Larrieu}}]{Thatte16}
{Thatte}, N.~A., {Clarke}, F., {Bryson}, I., {et~al.} 2016, in SPIE, Vol. 9908,
  Ground-based and Airborne Instrumentation for Astronomy VI, ed. C.~J.
  {Evans}, L.~{Simard}, \& H.~{Takami}, 99081X

\bibitem[{{Trump} {et~al.}(2006){Trump}, {Hall}, {Reichard}, {Richards},
  {Schneider}, {Vanden Berk}, {Knapp}, {Anderson}, {Fan}, {Brinkman},
  {Kleinman}, \& {Nitta}}]{Trump06}
{Trump}, J.~R., {Hall}, P.~B., {Reichard}, T.~A., {et~al.} 2006, \apjs, 165, 1

\bibitem[{{Urrutia} {et~al.}(2009){Urrutia}, {Becker}, {White}, {Glikman},
  {Lacy}, {Hodge}, \& {Gregg}}]{Urrutia09}
{Urrutia}, T., {Becker}, R.~H., {White}, R.~L., {et~al.} 2009, \apj, 698, 1095

\bibitem[{{van der Walt} {et~al.}(2011){van der Walt}, {Colbert}, \&
  {Varoquaux}}]{van11}
{van der Walt}, S., {Colbert}, S.~C., \& {Varoquaux}, G. 2011, Computing in
  Science and Engineering, 13, 22

\bibitem[{{Vanden Berk} {et~al.}(2001){Vanden Berk}, {Richards}, {Bauer},
  {Strauss}, {Schneider}, {Heckman}, {York}, {Hall}, {Fan}, {Knapp},
  {Anderson}, {Annis}, {Bahcall}, {Bernardi}, {Briggs}, {Brinkmann}, {Brunner},
  {Burles}, {Carey}, {Castander}, {Connolly}, {Crocker}, {Csabai}, {Doi},
  {Finkbeiner}, {Friedman}, {Frieman}, {Fukugita}, {Gunn}, {Hennessy},
  {Ivezi{\'c}}, {Kent}, {Kunszt}, {Lamb}, {Leger}, {Long}, {Loveday}, {Lupton},
  {Meiksin}, {Merelli}, {Munn}, {Newberg}, {Newcomb}, {Nichol}, {Owen}, {Pier},
  {Pope}, {Rockosi}, {Schlegel}, {Siegmund}, {Smee}, {Snir}, {Stoughton},
  {Stubbs}, {SubbaRao}, {Szalay}, {Szokoly}, {Tremonti}, {Uomoto}, {Waddell},
  {Yanny}, \& {Zheng}}]{vanden01}
{Vanden Berk}, D.~E., {Richards}, G.~T., {Bauer}, A., {et~al.} 2001, \aj, 122,
  549

\bibitem[{{Virtanen} {et~al.}(2020){Virtanen}, {Gommers}, {Burovski},
  {Oliphant}, {Weckesser}, {Cournapeau}, {Alexbrc}, {Peterson}, {Wilson},
  {Reddy}, {Mayorov}, {Endolith}, {Haberland}, {Nelson}, {Van Der Walt},
  {Laxalde}, {Brett}, {Polat}, {Larson}, {Millman}, {Lars}, {Van Mulbregt},
  {Eric-Jones}, {Carey}, {Moore}, {Kern}, {Leslie}, {Perktold}, {Striega}, \&
  {Feng}}]{Virtanen20}
{Virtanen}, P., {Gommers}, R., {Burovski}, E., {et~al.} 2020, {scipy/scipy:
  SciPy 1.5.2}

\bibitem[{{Voit} {et~al.}(1993){Voit}, {Weymann}, \& {Korista}}]{Voit93}
{Voit}, G.~M., {Weymann}, R.~J., \& {Korista}, K.~T. 1993, \apj, 413, 95

\bibitem[{{Wang} {et~al.}(2022){Wang}, {Luo}, {Brandt}, {Alexander}, {Bauer},
  {Gallagher}, {Huang}, {Liu}, \& {Stern}}]{Wang22}
{Wang}, C., {Luo}, B., {Brandt}, W.~N., {et~al.} 2022, \apj, 936, 95

\bibitem[{{W}es {M}c{K}inney(2010)}]{Mckinney10}
{W}es {M}c{K}inney. 2010, in {P}roceedings of the 9th {P}ython in {S}cience
  {C}onference, ed. {S}t\'efan van~der {W}alt \& {J}arrod {M}illman, 56 -- 61

\bibitem[{{Weymann} {et~al.}(1991){Weymann}, {Morris}, {Foltz}, \&
  {Hewett}}]{Wey91}
{Weymann}, R.~J., {Morris}, S.~L., {Foltz}, C.~B., \& {Hewett}, P.~C. 1991,
  \apj, 373, 23

\bibitem[{{Wright} {et~al.}(2010){Wright}, {Eisenhardt}, {Mainzer}, {Ressler},
  {Cutri}, {Jarrett}, {Kirkpatrick}, {Padgett}, {McMillan}, {Skrutskie},
  {Stanford}, {Cohen}, {Walker}, {Mather}, {Leisawitz}, {Gautier}, {McLean},
  {Benford}, {Lonsdale}, {Blain}, {Mendez}, {Irace}, {Duval}, {Liu}, {Royer},
  {Heinrichsen}, {Howard}, {Shannon}, {Kendall}, {Walsh}, {Larsen}, {Cardon},
  {Schick}, {Schwalm}, {Abid}, {Fabinsky}, {Naes}, \& {Tsai}}]{Wright10}
{Wright}, E.~L., {Eisenhardt}, P. R.~M., {Mainzer}, A.~K., {et~al.} 2010, \aj,
  140, 1868

\bibitem[{{Yang} {et~al.}(2018){Yang}, {Brandt}, {Vito}, {Chen}, {Trump},
  {Luo}, {Sun}, {Xue}, {Koekemoer}, {Schneider}, {Vignali}, \& {Wang}}]{Yang18}
{Yang}, G., {Brandt}, W.~N., {Vito}, F., {et~al.} 2018, \mnras, 475, 1887

\bibitem[{{Yong} {et~al.}(2018){Yong}, {King}, {Webster}, {Bate}, {O'Dowd}, \&
  {Labrie}}]{Yong18}
{Yong}, S.~Y., {King}, A.~L., {Webster}, R.~L., {et~al.} 2018, \mnras, 479,
  4153

\bibitem[{{York} {et~al.}(2000){York}, {Adelman}, {Anderson}, {Anderson},
  {Annis}, {Bahcall}, {Bakken}, {Barkhouser}, {Bastian}, {Berman}, {Boroski},
  {Bracker}, {Briegel}, {Briggs}, {Brinkmann}, {Brunner}, {Burles}, {Carey},
  {Carr}, {Castander}, {Chen}, {Colestock}, {Connolly}, {Crocker}, {Csabai},
  {Czarapata}, {Davis}, {Doi}, {Dombeck}, {Eisenstein}, {Ellman}, {Elms},
  {Evans}, {Fan}, {Federwitz}, {Fiscelli}, {Friedman}, {Frieman}, {Fukugita},
  {Gillespie}, {Gunn}, {Gurbani}, {de Haas}, {Haldeman}, {Harris}, {Hayes},
  {Heckman}, {Hennessy}, {Hindsley}, {Holm}, {Holmgren}, {Huang}, {Hull},
  {Husby}, {Ichikawa}, {Ichikawa}, {Ivezi{\'c}}, {Kent}, {Kim}, {Kinney},
  {Klaene}, {Kleinman}, {Kleinman}, {Knapp}, {Korienek}, {Kron}, {Kunszt},
  {Lamb}, {Lee}, {Leger}, {Limmongkol}, {Lindenmeyer}, {Long}, {Loomis},
  {Loveday}, {Lucinio}, {Lupton}, {MacKinnon}, {Mannery}, {Mantsch}, {Margon},
  {McGehee}, {McKay}, {Meiksin}, {Merelli}, {Monet}, {Munn}, {Narayanan},
  {Nash}, {Neilsen}, {Neswold}, {Newberg}, {Nichol}, {Nicinski}, {Nonino},
  {Okada}, {Okamura}, {Ostriker}, {Owen}, {Pauls}, {Peoples}, {Peterson},
  {Petravick}, {Pier}, {Pope}, {Pordes}, {Prosapio}, {Rechenmacher}, {Quinn},
  {Richards}, {Richmond}, {Rivetta}, {Rockosi}, {Ruthmansdorfer}, {Sandford},
  {Schlegel}, {Schneider}, {Sekiguchi}, {Sergey}, {Shimasaku}, {Siegmund},
  {Smee}, {Smith}, {Snedden}, {Stone}, {Stoughton}, {Strauss}, {Stubbs},
  {SubbaRao}, {Szalay}, {Szapudi}, {Szokoly}, {Thakar}, {Tremonti}, {Tucker},
  {Uomoto}, {Vanden Berk}, {Vogeley}, {Waddell}, {Wang}, {Watanabe},
  {Weinberg}, {Yanny}, {Yasuda}, \& {SDSS Collaboration}}]{York00}
{York}, D.~G., {Adelman}, J., {Anderson}, John~E., J., {et~al.} 2000, \aj, 120,
  1579

\bibitem[{{Yuan} \& {Wills}(2003)}]{Yuan03}
{Yuan}, M.~J., \& {Wills}, B.~J. 2003, \apjl, 593, L11

\bibitem[{{Zhou} {et~al.}(2006){Zhou}, {Wang}, {Wang}, {Wang}, {Yuan}, \&
  {Lu}}]{Zhou06}
{Zhou}, H., {Wang}, T., {Wang}, H., {et~al.} 2006, \apj, 639, 716

\end{thebibliography}

\begin{longrotatetable}
\begin{deluxetable}{lccccccccccccccc}
\tablenum{1}
\tablecaption{Optical Spectral Properties of 65 BAL Quasars from GNIRS-DQS \label{tab:table1}}
\tablecolumns{13}
\setlength{\tabcolsep}{2.8 pt}
\tablewidth{0.5 cm}
\tabletypesize{\scriptsize}
\tablehead{
\colhead{} & \colhead{} & \colhead{} & \colhead{Velocity Offset} & \colhead{EW$_{\rm H\beta}$} & \colhead{EW$_{\rm Fe~{II}}$} & \colhead {FWHM$_{\rm H\beta}$} & \colhead{EW$_{[\rm O~{III}]}$} & \colhead{FWHM$_{[\rm O~{III}]}$} & \colhead{Asymmetry$_{[\rm O~{III}]}$} & \colhead {log $\lambda L_{5100}$} & \colhead{log $M_{\rm BH}$} & \colhead{$L/L_{\rm Edd}$}\\
 \colhead{Quasar (SDSS J)} & \colhead{$z_{\rm H\beta}$}  & \colhead{$z_{[\rm O~{III}]}$} & \colhead{($\rm km~s^{-1}$)}& \colhead{({\AA})} & \colhead{({\AA})} & \colhead{($\rm km~s^{-1}$)} & \colhead{({\AA})} & \colhead{($\rm km~s^{-1}$)} & \colhead{} & \colhead{($\rm erg~s^{-1}$)} & \colhead{(\(M_\odot\))} & \colhead{} \
}

\startdata
$001249.89+285552.6$ & 3.248 & 3.237 & 807 & $47_{-14}^{+11}$ & $39\pm 3$ & $4622_{-442}^{+334}$ & $9_{-1}^{+1}$ & $2394_{-1}^{+1}$ & $0.015$ & 46.78 & 9.40 & 1.41 \\
$001355.10-012304.0$ & 3.391 & 3.380 & 793 & $97_{-45}^{+34}$ & $41.7\pm 0.3$ & $7323_{-1920}^{+1451}$ & $20_{-1}^{+1}$ & $1918_{-1}^{+1}$ & $-0.191$ & 46.60 & 9.87 & 0.33 \\
$004613.54+010425.7$ & 2.171 & 2.166 & 392 & $92_{-1}^{+1}$ & $32\pm 1$ & $3522_{-128}^{+97}$ & $24_{-1}^{+1}$ & $955_{-1}^{+1}$ & $-0.039$ & 46.48 & 9.21 & 1.14 \\
$013012.36+153157.9$ & 2.344 & 2.343 & 77 & $64_{-2}^{+1}$ & $18\pm 1$ & $4734_{-153}^{+116}$ & $15_{-1}^{+1}$ & $778_{-52}^{+40}$ & $-0.166$ & 46.35 & 9.43 & 0.51 \\
$013652.52+122501.5$ & 2.394 & 2.388 & 537 & $104_{-3}^{+2}$ & $43\pm 1$ & $4897_{-2116}^{+1599}$ & $30_{-1}^{+1}$ & $2041_{-1}^{+1}$ & $-0.097$ & 46.47 & 9.46 & 0.61 \\
$014018.20-013805.8$ & 2.232 & 2.235 & -319 & $51_{-1}^{+1}$ & $12.9\pm 0.3$ & $6212_{-171}^{+129}$ & $7_{-1}^{+1}$ & $1138_{-1}^{+1}$ & $-0.217$ & 46.64 & 9.80 & 0.41 \\
$014206.86+025713.0$ & 2.323 & 2.325 & -117 & $48_{-2}^{+2}$ & $23\pm 1$ & $4721_{-114}^{+86}$ & $17_{-1}^{+1}$ & $1496_{-139}^{+105}$ & $-0.077$ & 46.76 & 9.54 & 0.98 \\
$022007.64-010731.1$ & 3.436 & 3.429 & 469 & $85_{-3}^{+2}$ & $33.4\pm 0.1$ & $3867_{-154}^{+116}$ & $19_{-1}^{+1}$ & $2819_{-1}^{+1}$ & $0.108$ & 46.62 & 9.33 & 1.16\\
$025042.45+003536.7$ & 2.398 & \nodata & \nodata & $122_{-3}^{+2}$ & $28\pm 1$ & $8886_{-763}^{+577}$ & ${< 1}$ & \nodata & \nodata & 46.45 & 10.04 & 0.16\\
$073519.68+240104.6$ & 3.295 & \nodata & \nodata & $62_{-4}^{+3}$ & $32\pm 1$ & $5227_{-256}^{+193}$ & ${< 1}$ & \nodata & \nodata & 46.51 & 9.50 & 0.61\\
$080636.81+345048.5$ & 1.548 & 1.547 & 121 & $87_{-39}^{+29}$ & $59\pm 3$ & $4427_{-863}^{+653}$ & $12_{-2}^{+2}$ & $2936_{-501}^{+379}$ & $0.413$ & 46.11 & 9.12 & 0.61 \\
$081114.66+172057.4$ & 2.341 & 2.341 & -2 & $59_{-1}^{+1}$ & $9.8\pm 0.5$ & $6841_{-196}^{+149}$ & $17_{-1}^{+1}$ & $1922_{-1}^{+1}$ & $-0.436$ & 46.49 & 9.85 & 0.26 \\
$083745.74+052109.4$ & 2.364 & 2.361 & 306 & $71_{-4}^{+3}$ & $30\pm 1$ & $3950_{-165}^{+125}$ & $22_{-1}^{+1}$ & $2275_{-1}^{+1}$ & $-0.149$ & 46.31 & 9.20 & 0.78\\
$084133.15+200525.7$ & 2.360 & \nodata & \nodata & $27_{-1}^{+1}$ & $28\pm 1$ & $5582_{-189}^{+143}$ & ${< 1}$ & \nodata & \nodata & 47.00 & 9.60 & 1.49\\
$084401.95+050357.9$ & 3.358 & 3.341 & 1220 & $50_{-1}^{+1}$ & $19\pm 1$ & $5964_{-1}^{+1}$ & $43_{-1}^{+1}$ & $5811_{-1}^{+1}$ & $-0.042$ & 47.19 & 9.97 & 0.97\\
$084729.52+441616.7$ & 2.346 & 2.343 & 261 & $71_{-13}^{+10}$ & $30.7\pm 0.1$ & $4257_{-479}^{+362}$ & $15_{-1}^{+1}$ & $719_{-1}^{+1}$ & $0.143$ & 46.37 & 9.29 & 0.73\\
$085046.17+522057.4$ & 2.235 & \nodata & \nodata & $59_{-1}^{+1}$ & $36\pm 2$ & $4445_{-298}^{+226}$ & ${< 1}$ & \nodata & \nodata & 46.62 & 9.38 & 1.05 \\ 
$091301.01+422344.7$ & 2.317 & 2.314 & 309 & $81_{-1}^{+1}$ & $38\pm 1$ & $4512_{-150}^{+114}$ & $41_{-1}^{+1}$ & $1795_{-1}^{+1}$ & $-0.497$ & 46.50 & 9.39 & 0.78 \\
$091328.23+394443.9^{*}$ & 1.582 & 1.587 & -605 & $39_{-4}^{+3}$ & $45\pm 4$ & $2543_{-406}^{+307}$ & $11_{-1}^{+1}$ & $1289_{-1}^{+1}$ & $-0.019$ & 46.02 & 8.44 & 2.41 \\
$091425.72+504854.9$ & 2.367 & 2.368 & -116 & $64_{-4}^{+3}$ & $45\pm 1$ & $4584_{-123}^{+93}$ & $8_{-1}^{+1}$ & $1258_{-167}^{+126}$ & $-0.384$ & 46.32 & 9.24 & 0.74 \\
$091716.79+461435.3$ & 1.630 & 1.625 & 548 & $19_{-4}^{+3}$ & ${1 \pm 1 \times 10^{-7}}$ & $5571_{-2076}^{+1569}$ & $31_{-1}^{+1}$ & $778_{-1}^{+1}$ & $-0.144$ & 46.31 & 9.65 & 0.28 \\
$093251.98+023727.0$ & 2.172 & 2.171 & 131 & $51_{-1}^{+1}$ & $38\pm 1$ & $5988_{-150}^{+113}$ & $10_{-1}^{+1}$ & $1917_{-1}^{+1}$ & $0.026$ & 46.47 & 9.52 & 0.54\\
$094328.94+140415.6^{*}$ & 2.412 & \nodata & \nodata & $53_{-6}^{+5}$ & $44\pm 1$ & $5940_{-846}^{+640}$ & ${< 1}$ & \nodata & \nodata & 46.55 & 9.52 & 0.65 \\
$094427.27+614424.6$ & 2.338 & 2.340 & -169 & $79_{-3}^{+2}$ & $24\pm 1$ & $4898_{-3015}^{+2279}$ & $38_{-2}^{+2}$ & $1795_{-246}^{+186}$ & $-0.229$ & 46.41 & 9.47 & 0.53 \\
$094902.38+531241.5$ & 1.613 & 1.609 & 430 & $64_{-2}^{+1}$ & $48\pm 1$ & $4345_{-161}^{+121}$ & $4_{-1}^{+1}$ & $1139_{-1}^{+1}$ & $0.915$ & 46.19 & 9.11 & 0.74 \\
$095746.75+565800.7$ & 1.585 & 1.576 & 1103 & $26_{-5}^{+4}$ & $61\pm 1$ & $2613_{-481}^{+364}$ & $2_{-1}^{+1}$ & $321_{-1}^{+1}$ & $-0.046$ & 46.39 & 8.20 & 9.36 \\
$100610.55+370513.8$ & 3.195 & 3.202 & -490 & $72_{-4}^{+3}$ & $51.2\pm 1 \times10^{-16}$ & $5601_{-724}^{+547}$ & $20_{-1}^{+1}$ & $674_{-1}^{+1}$ &  $-0.238$ & 46.78 & 9.61 & 0.87 \\
$100653.26+011938.7$ & 2.298 & 2.303 & -427 & $62_{-2}^{+1}$ & $46\pm 1$ & $5850_{-400}^{+302}$ & $21_{-1}^{+1}$ & $1735_{-1}^{+1}$ & $-0.408$ & 46.40 & 9.47 & 0.52 \\
$101542.04+430455.6$ & 2.428 & 2.415 & 1115 & $59_{-5}^{+4}$ & $54\pm 2$ & $3763_{-181}^{+137}$ & $5_{-1}^{+1}$ & $2875_{-256}^{+194}$ & $-0.021$ & 46.48 & 9.07 & 1.57 \\
$102154.00+051646.3$ & 3.448 & \nodata & \nodata & $61_{-1}^{+1}$ & $16.5\pm 0.1$ & $5254_{-300}^{+226}$ & ${< 1}$ & \nodata & \nodata & 46.70 & 9.68 & 0.63 \\
$103246.19+323618.0$ & 2.380 & \nodata & \nodata & $36_{-1}^{+1}$ & $37\pm 1$ & $4755_{-330}^{+249}$ & ${< 1}$ & \nodata & \nodata & 46.44 & 9.21 & 1.04 \\
$103405.73+463545.4$ & 2.213 & 2.215 & -203 & $63_{-3}^{+2}$ & $26\pm 1$ & $6489_{-254}^{+191}$ & $17_{-11}^{+12}$ & $718_{-718}^{+974}$ & $-0.665$ & 46.44 & 9.69 & 0.34 \\
$103718.23+302509.1$ & 2.299 & 2.288 & 977 & $100_{-76}^{+57}$ & $40.1\pm 1 \times10^{-10}$ & $4724_{-163}^{+123}$ & $26_{-1}^{+1}$ & $3538_{-1}^{+1}$ & $0.033$ & 46.39 & 9.40 & 0.60\\
$104621.57+483322.7$ & 1.580 & \nodata & \nodata & $39_{-2}^{+1}$ & $60\pm 3$ & $4222_{-552}^{+417}$ & ${< 1}$ & \nodata & \nodata & 46.15 & 8.79 & 1.41\\
$104941.58+522348.9$ & 2.367 & 2.369 & -115& $97_{-3}^{+2}$ & $21\pm 1$ & $5663_{-517}^{+391}$ & $40_{-8}^{+6}$ & $2152_{-787}^{+595}$ & $-0.198$ & 46.27 & 9.57 & 0.31 \\
$110148.85+054815.5$ & 1.589 & 1.584 & 600 & $48_{-2}^{+1}$ & $17\pm 1$ & $7944_{-605}^{+457}$ & $17_{-1}^{+1}$ & $3473_{-329}^{+249}$ & $0.00022$ & 46.38 & 9.86 & 0.20 \\
$111313.29+102212.4$ & 2.261 & \nodata & \nodata & $81_{-1}^{+1}$ & $42\pm 1$ & $4752_{-190}^{+144}$ & ${< 1}$ & \nodata & \nodata & 46.58 & 9.45 & 0.81 \\
$111352.53+104041.9$ & 1.607 & 1.609 & -233 & $26_{-2}^{+1}$ & $5.4\pm 0.1$ & $6068_{-1005}^{+760}$ & $5_{-1}^{+1}$ & $1675_{-1}^{+1}$ & $-0.554$ & 46.39 & 9.69 & 0.31 \\
$111920.98+232539.4$ & 2.288 & 2.289 & -53 & $46_{-1}^{+1}$ & $9.5\pm 1 \times10^{-15}$ & $3901_{-167}^{+126}$ & $28_{-1}^{+1}$ & $1557_{-1}^{+1}$ &  $-0.180$ & 46.35 & 9.28 & 0.71 \\
$112127.79+254758.9$ & 1.602 & 1.602 & -8 & $77_{-2}^{+1}$ & $4.4\pm 0.2$ & $4700_{-2105}^{+1591}$ & $39_{-2}^{+1}$ & $898_{-102}^{+77}$ & $-0.395$ & 46.39 & 9.52 & 0.46 \\
$112938.46+440325.0$ & 2.212 & 2.212 & -1 & $128_{-5}^{+4}$ & $7.1\pm 0.1$ & $3267_{-154}^{+117}$& $46_{-5}^{+4}$ & $659_{-156}^{+118}$ & $-0.493$ & 46.62 & 9.30 & 1.24 \\
$113048.45+225206.6$ & 2.378 & 2.364 & 1285 & $34_{-15}^{+11}$ & $9.9\pm 1 \times10^{-15}$ & $5358_{-282}^{+213}$ & $11_{-11}^{+19}$ & $3119_{-3119}^{+7805}$ & $0.0067$ & 46.39 & 9.55 & 0.42 \\
$113330.17+144758.8$ & 3.254 & 3.238 & 1158 & $68_{-4}^{+3}$ & $35\pm 2$ & $4629_{-249}^{+188}$ & $9_{-2}^{+2}$ & $2158_{-1031}^{+779}$ & $-0.00031$ & 46.65 & 9.46 & 0.93 \\
$113740.61+630256.9$ & 2.318 & 2.323 & -493 & $59_{-2}^{+2}$ & $23\pm 1$ & $7995_{-411}^{+311}$ & $4_{-1}^{+1}$ & $757_{-1}^{+1}$ & $-0.578$ & 46.41 & 9.87 & 0.21 \\
$113924.64+332436.9$ & 2.310 & 2.315 & -396 & $41_{-2}^{+1}$ & $60\pm 2$ & $5454_{-574}^{+434}$ & $1.5_{-1}^{+1}$ & $1672_{-1}^{+1}$ & $-0.00044$ & 46.31 & 9.12 & 0.95\\
$114323.71+193448.0$ & 3.358 & 3.350 & 554 & $77_{-1}^{+1}$ & $18\pm 1$ & $5003_{-140}^{+106}$ & $40_{-1}^{+1}$ & $2750_{-1}^{+1}$ & $-0.202$ & 46.78 & 9.69 & 0.74\\
$114705.24+083900.6$ & 1.602 & 1.603 & -122 & $51_{-2}^{+2}$ & $4\pm 1$ & $4778_{-501}^{+378}$ & $36_{-1}^{+1}$ & $959_{-62}^{+47}$ & $-0.345$ & 46.51 & 9.58 & 0.51\\
$114738.35+301717.5$ & 3.358 & \nodata & \nodata & $46_{-4}^{+3}$ & $29\pm 1$ & $4430_{-707}^{+534}$ & ${< 1}$ & \nodata & \nodata & 46.74 & 9.42 & 1.23\\
$115747.99+272459.6$ & 2.230 & 2.216 & 1276 & $67_{-3}^{+2}$ & $37\pm 2$ & $3896_{-314}^{+238}$ & $12_{-1}^{+1}$ & $1288_{-1}^{+1}$ & $-0.00079$ & 46.63 & 9.29 & 1.33\\
$133342.56+123352.7$ & 3.281 & 3.258 & 1650 & $59_{-1}^{+1}$ & $30\pm 1$ & $4150_{-190}^{+143}$ & $11_{-1}^{+1}$ & $2645_{-1}^{+1}$ & $-0.022$ & 46.79 & 9.43 & 1.36\\
$140058.79+260619.4$ & 2.366 & 2.363 & 238 & $54_{-1}^{+1}$ & $31\pm 1$ & $2898_{-132}^{+100}$ & $10_{-1}^{+1}$ & $1736_{-1}^{+1}$ & $-0.450$ & 46.40 & 8.92 & 1.85 \\
$141321.05+092204.8$ & 3.324 & 3.313 & 737 & $69_{-2}^{+1}$ & $22\pm 1$ & $4019_{-115}^{+87}$ & $12_{-1}^{+1}$ & $2097_{-184}^{+139}$ & $-0.017$ & 46.90 & 9.52 & 1.43\\
$142013.03+253403.9$ & 2.234 & 2.235 & -130 & $36_{-3}^{+2}$ & $47\pm 2$ & $2820_{-228}^{+173}$ & $6_{-5}^{+6}$ & $539_{-539}^{+9151}$ & $0.048$ & 46.56 & 8.71 & 4.29\\
$142500.24+494729.2$ & 2.261 & \nodata & \nodata & $58_{-2}^{+1}$ & $45\pm 2$ & $4134_{-169}^{+128}$ & ${< 1}$ & \nodata & \nodata & 46.46 & 9.19 & 1.14\\
$150205.58-024038.5$ & 2.210 & 2.192 & 1667 & $52_{-2}^{+1}$ & $31\pm 1$ & $4459_{-217}^{+164}$ & $21_{-1}^{+1}$ & $1994_{-1}^{+1}$ & $0.038$ & 46.48 & 9.32 & 0.87\\
$151123.30+495101.2$ & 2.396 & \nodata & \nodata & $47_{-1}^{+1}$ & $55\pm 1$ & $6907_{-306}^{+231}$ & ${< 1}$ & \nodata & \nodata & 46.67 & 9.59 & 0.73\\
$151341.89+463002.7^{*}$ & 1.574 & 1.572 & 203 & $51_{-2}^{+1}$ & $33\pm 1$ & $3937_{-83}^{+63}$ & $13_{-1}^{+1}$ & $1257_{-1}^{+1}$ & $-0.486$ & 46.42 & 9.10 & 0.87\\
$153248.95+173900.8$ & 2.348 & 2.350 & -181 & $55_{-1}^{+1}$ & $23\pm 1$ & $7286_{-434}^{+328}$ & $12_{-1}^{+1}$ & $1377_{-1277}^{+7381}$ & $-0.0089$ & 46.56 & 9.78 & 0.26\\
$154550.37+554346.2$ & 2.164 & 2.159 & 432 & $111_{-5}^{+4}$ & $6.5\pm 1 \times10^{-11}$ & $3152_{-1}^{+1}$ & $42_{-1}^{+1}$ & $1137_{-1}^{+1}$ & $-0.327$ & 46.72 & 9.25 & 1.24\\
$160207.67+380743.0$ & 1.592 & 1.581 & 1214 & $65_{-17}^{+13}$ & $29\pm 1$ & $7142_{-2455}^{+1856}$ & $24_{-1}^{+1}$ & $2340_{-122}^{+92}$ & $0.541$ & 46.72 & 9.89 & 0.40\\
$160552.97+292141.4$ & 2.327 & \nodata & \nodata & $87_{-4}^{+3}$ & $20.6\pm 1 \times10^{-11}$ & $4345_{-224}^{+169}$ & ${< 1}$ & \nodata & \nodata & 46.56 & 9.46 & 0.76\\
$163125.10+174810.0$ & 2.185 & 2.180 & 436 & $76_{-2}^{+2}$ & $7.5\pm 0.3$ & $5617_{-313}^{+237}$ & $33_{-2}^{+2}$ & $1736_{-259}^{+196}$ & $-0.106$ & 46.68 & 9.79 & 0.47\\ 
$220344.98+235729.3^{*}$ & 2.159 & 2.158 & 79 & $76_{-1}^{+1}$ & $14\pm 1$ & $6974_{-358}^{+271}$ & $15_{-2}^{+1}$ & $1436_{-521}^{+690}$ & $-0.164$ & 46.35 & 9.80 & 0.33\\
$223934.45-004707.2$ & 2.226 & 2.215 & 1100 & $62_{-2}^{+1}$ & $38\pm 2$ & $6179_{-247}^{+187}$ & $11_{-1}^{+1}$ & $1681_{-1}^{+1}$ & $0.103$ & 46.49 & 9.60 & 0.47\\
$225608.48+010557.8$ & 2.267 & 2.263 & 350 & $61_{-4}^{+3}$ & $33.1\pm 1 \times10^{-10}$ & $2830_{-363}^{+275}$ & $12_{-1}^{+1}$ & $3828_{-1}^{+1}$ & $-0.235$ & 46.41 & 8.92 & 1.90 \\
\enddata
\tablenotetext{*}{LoBAL quasars identified by \citet{Trump06} or through visual inspection.}
\end{deluxetable}
\end{longrotatetable}

\startlongtable
\begin{deluxetable*}{lccccccccccccccc}
\tablenum{2}
\tablecaption{UV Spectral Properties of 65 BAL Quasars from GNIRS-DQS \label{tab:table2}}
\tablecolumns{17}
\setlength{\tabcolsep}{2.7 pt}
\tablewidth{0.5 cm}
\tabletypesize{\scriptsize}
\tablehead{
\colhead{} & \colhead{} & \colhead{} & \colhead{Velocity Offset} & \colhead{EW$_{\rm C~{IV}}$$^{b}$} & \colhead {Blueshift$_{\rm C~{IV}}$$^{b}$} & \colhead {BI} & \colhead{AI} & \colhead{BI/AI}\\
 \colhead{Quasar (SDSS J)} & \colhead{$z_{\rm vi^{c}}$}  & \colhead{$z_{\rm sys}$} & \colhead{($\rm km~s^{-1}$)} & \colhead{({\AA})} & \colhead{($\rm km~s^{-1}$)} & \colhead{($\rm km~s^{-1}$)} & \colhead{($\rm km~s^{-1}$)} & \colhead{Ref.$^{a}$}\
}

\startdata
$001249.89+285552.6$ & 3.236 & 3.233 & 213 & 25 & 1582 & 164 & 1924 & 1\\
$001355.10-012304.0$ & \nodata & 3.380 & \nodata & 30 & 1370 & 1226 & 1598 & 3\\
$004613.54+010425.7$ & 2.150 & 2.165 & -1422 & 36 & 2264 & 3393 & 7033 & 1\\
$013012.36+153157.9$ & 2.349 & 2.343 & 538  & 15 & 435 & 5 & 2468 & 1\\
$013652.52+122501.5$ & 2.393 & 2.388 & 443 & 68 & 1060 & 5177 & 6130 & 1\\
$014018.20-013805.8$ & 2.235 & 2.236 & -93 & 33 & 1522 & 1091 & 1404 & 1\\
$014206.86+025713.0$ & 2.315 & 2.323 & -722 & 19 & 640 & 40 & 1229 & 3\\
$022007.64-010731.1$ & 3.441 & 3.428 & 881 & 35 & 2127 & 5649 & 8132 & 1\\
$025042.45+003536.7$ & 2.387 & 2.398 & -971 & 42 & 978 & 5336 & 7361 & 1\\
$073519.68+240104.6$ & 3.278 & 3.282 & -280 & 26 & 1865 & 25 & 1983 & 1\\
$080636.81+345048.5$ & \nodata & 1.549 & \nodata & \nodata & \nodata & 0 & 0 & 5\\
$081114.66+172057.4$ & 2.323 & 2.341 & -1616 & 26 & 1427 & 405 & 2687 & 1\\
$083745.74+052109.4$ & 2.355 & 2.362 & -625 & 42 & 1582 & 385 & 3137 & 1\\
$084133.15+200525.7$ & 2.342 & 2.356 & -1252 & 9 & 6936 & 7235 & 1178 & 1\\
$084401.95+050357.9$ & \nodata & 3.340 & \nodata  & 16 & 4458 & 8972 & 9558 & 3\\
$084729.52+441616.7$ & 2.347 & 2.344 & 269 & 16 & 1274 & 572 & 1785 & 1\\
$085046.17+522057.4$ & 2.230 & 2.234 & -371 & \nodata & \nodata & 1065 & 1641 & 3\\ 
$091301.01+422344.7$ & 2.315 & 2.314 & 91 & 59 & 1563 & 645 & 980 & 1\\
$091328.23+394443.9$ & \nodata & 1.587 & \nodata & \nodata & \nodata & 390 & 508 & 3\\
$091425.72+504854.9$ & 2.345 & 2.369 & -2493 & 28 & 1171 & 3817 & 7286 & 1\\
$091716.79+461435.3$ & \nodata & 1.625 & \nodata & \nodata & \nodata & 0 & 216 & 3\\
$093251.98+023727.0$ & 2.165 & 2.170 & -473 & 37 & 1928 & 4436 & 6935 & 1\\
$094328.94+140415.6$ & 2.430 & 2.408 & -704 & \nodata & \nodata & 0 & 1402 & 3\\
$094427.27+614424.6$ & 2.333 & 2.338 & -449 & 67 & 703 & 4169 & 6845 & 1\\
$094902.38+531241.5$ & 1.611 & 1.609 & 230 & 49 & 2502 & 753 & 2368 & 1\\
$095746.75+565800.7$ & 1.575 & 1.576 & -117 & 45 & 2925 & 808 & 0 & 2\\
$100610.55+370513.8$ & 3.204 & 3.202 & 143 & 53 & 1607 & 1441 & 1567 & 1\\
$100653.26+011938.7$ & 2.298 & 2.303 & -454 & 37 & 2382 & 6190 & 6715 & 1\\
$101542.04+430455.6$ & 2.364 & 2.417 & 702 & 13 & 4704 & 383 & 2079 & 1\\
$102154.00+051646.3$ & 3.439 & 3.448 & -607 & 24 &2853 & 9177 & 1230 & 1\\
$103246.19+323618.0$ & 2.380 & 2.379 & 89 & 11 & 5562 & 4393 & 5752 & 3\\
$103405.73+463545.4$ & 2.215 & 2.215 & $< |90|$ & 32 & 1816 & 0 & 699 & 1\\
$103718.23+302509.1$ & 2.293 & 2.288 & 456 & 44 & 1315 & 252 & 931 & 1\\
$104621.57+483322.7$ & \nodata & 1.580 & \nodata & \nodata & \nodata & 1676 & 2547 & 3\\
$104941.58+522348.9$ & \nodata & 2.364 & \nodata & 50 & 736 & 934 & 3557 & 3\\
$110148.85+054815.5$ & \nodata & 1.584 & \nodata & \nodata & \nodata & 0 & 1709 & 3\\
$111313.29+102212.4$ & 2.261 & 2.259 & 184 & 28 & 2349 & 613 & 1381 & 1\\
$111352.53+104041.9$ & \nodata & 1.609 & \nodata & \nodata & \nodata & 0 & 1939 & 3\\
$111920.98+232539.4$ & 2.289 & 2.289 & $< |90|$ & 57 & 547 & 220 & 1907 & 3\\
$112127.79+254758.9$ & 1.587 & 1.601 & -1615 & 67 & 523 & 1008 & 3251 & 1\\
$112938.46+440325.0$ & 2.210 & 2.212 & 93 & \nodata & \nodata & 0 & 605 & 3\\
$113048.45+225206.6$ & 2.370 & 2.364 & 535 & 42 & 1782 & 6596 & 7168 & 1\\
$113330.17+144758.8$ & 3.252 & 3.242 & 424 & \nodata & \nodata & 616 & 1940 & 3\\
$113740.61+630256.9$ & 2.322 & 2.323 & -90 & 52 & 1408 & 5169 & 5889 & 3\\
$113924.64+332436.9$ & 2.314 & 2.315 & -91 & 28 & 3626 & 4283 & 5198 & 1\\
$114323.71+193448.0$ & 3.348 & 3.350 & -138 & 43 & 1308 & 0 & 1470 & 1\\
$114705.24+083900.6$ & 1.604 & 1.601 & 346 & \nodata & \nodata & 0 & 242 & 3\\
$114738.35+301717.5$ & 3.353 & 3.358 & -344 & 7 & 1574 & 0 & 336 & 1\\
$115747.99+272459.6$ & 2.206 & 2.216 & -933 & 32 & 1322 & 4076 & 5548 & 1\\
$133342.56+123352.7$ & 3.275 & 3.258 & 1198 & 20 & 2466 & 0 & 514 & 1\\
$140058.79+260619.4$ & 2.351 & 2.363 & -1071 & 37 & 1992 & 1851 & 2432 & 1\\
$141321.05+092204.8$ & 3.327 & 3.324 & 765 & 20 & 1734 & 58 & 964 & 1\\
$142013.03+253403.9$ & 2.235 & 2.235 & $< |90|$ & 15 & 5255 & 2469 & 4993 & 1\\
$142500.24+494729.2$ & 2.260 & 2.261 & -92 & 15 & 3953 & 2648 & 3156 & 1\\
$150205.58-024038.5$ & \nodata & 2.192 & \nodata & 31 & 2774 & 6929 & 8100 & 3\\
$151123.30+495101.2$ & 2.380 & 2.379 & 265 & 19 & 4131 & 178 & 1954 & 1\\
$151341.89+463002.7$ & \nodata & 1.572 & \nodata & \nodata & \nodata & 2661 & 4651 & 3\\
$153248.95+173900.8$ & 2.350 & 2.350 & $< |90|$ & 31 & 436 & 1437 & 4057 & 1\\
$154550.37+554346.2$ & 2.158 & 2.159 & -95 & 43 & 1050 & 0 & 350 & 3\\
$160207.67+380743.0$ & \nodata & 1.581 & \nodata  & \nodata & \nodata & 0 & 0 & 4\\
$160552.97+292141.4$ & 2.321 & 2.333 & -1080 & 53 & 678 & 1767 & 3858 & 1\\
$163125.10+174810.0$ & 2.180 & 2.180 & $< |90|$ & 20 & 1708 & 3042 & 5239 & 1\\ 
$220344.98+235729.3$ & \nodata & 2.157 & \nodata & 31 & -81 & 2170 & 3318 & 3\\
$223934.45-004707.2$ & 2.221 & 2.215 & 560 & 39 & 1040 & 110 & 349 & 1\\
$225608.48+010557.8$ & 2.268 & 2.263 & 460 & 40 & 1814 & 1217 & 3993 & 1\\
\enddata
\tablenotetext{a}{References for the Balnicity Index and Absorption Index, Column (13): (1) \citet{Paris17}; (2) \citet{Paris18}; (3) \citet{Lyke20}; (4) \citet{M21}; (5) \citet{Abazajian09}}
\tablenotetext{b}{Data taken from R20 for 52 of GNIRS-DQS BAL quasars that appear in their sample.}
\tablenotetext{c}{Value based on the visually inspected redshift measurement in SDSS DR16 \citep[Table D1, Column 17]{Lyke20} for 52 BAL quasars.}
\end{deluxetable*}

\begin{longrotatetable}
\begin{deluxetable*}{lccccccccccccc}
\tablenum{3} 
\tablecaption{BAL and non-BAL Quasars Property Statistics and Comparisons \label{tab:table3}}
\tablecolumns{13}
\tablewidth{0pt}
\tabletypesize{\scriptsize}
\tablehead{
\colhead{Property} & \colhead{$\mu_{\rm BAL}$} & \colhead{$\mu_{\rm non-BAL}$} & \colhead{$\rm SEM_{\rm BAL}$} & \colhead {$\rm SEM_{\rm non-BAL}$} & \colhead{$\sigma_{\rm BAL}$} & \colhead{$\sigma_{\rm non-BAL}$} & \colhead{$\rm Med_{\rm BAL}$} & \colhead{$\rm Med_{\rm non-BAL}$} & \colhead{A-D $p$-value} & \colhead{K-S $p$-value}\
}

\startdata
log EW$_{\rm H\beta}$ & 1.77 & 1.77 & 0.02 & 0.01 & 0.16 & 0.14 & 1.79 $\pm~0.36$ & 1.79 $\pm~0.24$ & 0.25 & 0.84\\
log EW$_{[\rm O~{III}]}$ & 1.19 & 1.14 & 0.04 & 0.03 & 0.33 & 0.35 & 1.22 $\pm~0.23$ & 1.17 $\pm~0.17$ & 0.25 & 0.69\\
log FWHM$_{\rm H\beta}$ & 3.68 & 3.63 & 0.02 & 0.01 & 0.12 & 0.13 & 3.68 $\pm~2.21$ & 3.62  $\pm~1.98$ & \textbf{0.002} & 0.11\\
log FWHM$_{[\rm O~{III}]}$ & 3.19 & 3.14 & 0.03 & 0.02 & 0.24 & 0.26 & 3.23 $\pm~0.04$ & 3.22 $\pm~0.04$ & 0.25 & 0.69\\
Asymmetry$_{[\rm O~{III}]}$ & 
-0.14 & -0.20 & 0.04 & 0.03 & 0.29 & 0.32 & -0.13 $\pm~0.06$ & -0.20 $\pm~0.03$ & 0.23 & 0.14\\
$R_{\rm Fe~{II}}$ & 0.55 & 0.54 & 0.05 & 0.03 & 0.40 & 0.39 & 0.44 $\pm~0.05$ & 0.48$\pm~0.04$ & 0.25 & 0.86\\
log $M_{\rm BH}$ & 9.42 & 9.30 & 0.04 & 0.02 & 0.34 & 0.33 & 9.46 $\pm~0.04$ & 9.29 $\pm~0.03$ & \textbf{0.004} & 0.19\\ 
$L/L_{\rm Edd}$ & 1.00 & 1.23 & 0.15 & 0.11 & 1.22 & 1.50 & 0.74 $\pm~0.06$ & 1.04 $\pm~0.08$ & 0.02 & 0.18\\
Velocity Offset ${[\rm O~{III}]}$ ($\rm km~s^{-1}$) & 353.85 & 325.16 & 79.92 & 38.26 & 576.38 & 495.86 & 283.99$\pm~112.77$  & 308.38 $\pm~35.10$& 0.18 & 0.29\\
Velocity Offset $(z_{\rm vi}-z_{\rm sys})$ ($\rm km~s^{-1}$) & -224.41 & -226.60 & 133.06 & 68.86 & 959.50 & 840.56 & -90.39 $\pm~72.70$ & -92.19 $\pm~61.71$ & 0.25 & 0.98\\
\enddata
\tablecomments{A comparison of several BAL and non-BAL spectral properties with probabilities of test statistics (see Section \ref{sec:results}). The $p$-values highlighted in bold indicate rejection of null hypothesis at 95\% and 99\% significance level.}
\end{deluxetable*}
\end{longrotatetable}

\end{document}